\documentstyle[preprint,prb,aps]{revtex}
\input{epsf}
\def\k{{\bf k}}
\begin{document}
\draft
\title{Implementation of an all-electron GW Approximation  using the Projector 
Augmented Wave method: I. Formulation and application to the electronic 
structure of semiconductors} 
\author{B.~Arnaud and  M.~Alouani}
\address{Institut de Physique et de Chimie des Mat\'eriaux de
Strasbourg (IPCMS), 23 rue du Loess,67037
Strasbourg, France}
\date{\today}
\maketitle

\begin{abstract}
We have implemented the so called $GW$ approximation (GWA)
based on  an all-electron full-potential Projector Augmented Wave (PAW)
method. For the  screening of the Coulomb interaction $W$ we tested three 
different plasmon-pole dielectric function 
 models, and showed that the accuracy of the quasiparticle energies is 
$not$ sensitive to the the details of  these models. 
We have then applied this new method to compute the quasiparticle band 
structure of some small, medium and large-band-gap semiconductors: Si, 
GaAs, AlAs, InP,  SiMg$_2$, C and  (insulator) LiCl. 
A special attention was devoted to the convergence of the self-energy with
respect to both the {\bf k}-points in the Brillouin zone and to the number
of reciprocal space $\bf G$-vectors. The most important result is that 
although the all-electron GWA improves considerably the quasiparticle
band structure of semiconductors, it does $not$ always provide the correct 
energy band gaps as originally claimed by GWA pseudopotential type of
calculations. We argue that  the  decoupling between the  valence and 
core electrons is a problem, and is some what hidden in a  
pseudopotential type of approach. 
\end{abstract}

\section{Introduction}
In the last few years, with the increase of  computer power, many 
researchers have tried to unify two
major but  different approaches  to the computation of electronic
properties of materials: (1) The local density approximation (LDA) to
the Kohn-Sham equations\cite{Hohenberg} 
where the correlation effects are
included in an average way through the use of the parameterized local
exchange-correlation potential.\cite{exc}  (2) Many-body models based
on simplified and parameterized Hamiltonians omitting much of the
subtlety of the chemical bonding but handling the correlation effects
effectively.  The aim of the unification being to provide a good
description of the quasi-particles properties in {\it real} solids
without omitting either the subtlety of chemical bonding or the
correlated nature of the 
electrons.\cite{Hedin1,Hedin2,Hybertsen1,Godby1,Godby2,Godby3,vonderlinden,Horsch,ummels,bobbert,Hamada,Hott,arya,shirleylouie,shirleyzhu,review_gw1,review_gw2,kulikov,steiner}

That LDA is inadequate has been known for example in the case of
semiconductors:  the  LDA  band gaps are  at least 50$\%$ smaller than
experimental values. It was shown by Perdew and 
Levy\cite{perdew} and
Sham and Schl\"uter\cite{shamschlueter} that the difference between 
the highest
occupied and lowest unoccupied state of the density functional theory
(DFT) eigenvalues is not the true quasiparticle band gap, but differs
from it by $\Delta$, the discontinuity in the exchange-correlation
potential when an electron is added to the system. Later, it was shown
that quasiparticle calculation based on the 
the GW approximation (GWA) of Hedin\cite{Hedin1,Hedin2}  produced band gaps
which were in good agreement with experiment.
\cite{Hybertsen1,Godby1,Godby2,Godby3,vonderlinden,Horsch,ummels,bobbert,Hamada,Hott,arya,shirleylouie,shirleyzhu,review_gw1,review_gw2}
This development in  the theoretical study of the electronic
structure of solids is the inclusion of many-body effects in the
calculation,  principally, through the computation of the quasiparticle 
self-energy and the results are an adjustment of  the energy splittings 
obtained within the local density approximation.\cite{Hohenberg}

After this initial success  the GWA has been extensively
used to study other properties of different type of materials:  (1) The
band-width narrowing in alkali-metals, and their 
clusters.\cite{northruphl,saito}
(2) The surfaces states of semiconductors, i.e., improving band gaps of
surface states,\cite{zhu} looking for dimer buckling in Si 
surfaces.\cite{northrup}
(3) The effects of correlation on the valence off-set between different
bulk semiconductors.\cite{zhangclth} (4) The character of the band gaps of
superlattices and anisotropy of optical matrix elements.\cite{zhanghclt}  
(5) The orientational disorder and photoemission spectra of solid 
C$_{60}$.\cite{shirleylouie}
(6) The electronic properties of elemental Ni and its energy
loss spectra.\cite{arya} (7) The quasiparticle properties of atoms using
various GWA.\cite{shirleymartin} (8) The Schottky barrier between
a metal and a semiconductor.\cite{charlesworth} (9) The inclusion of excitonic effects
in the calculation of the dielectric function of semiconductors \cite{albrecht,benedict,rohlfing1}. 
All these different studies in
a very short time established the GWA as a good
`first-principles' method for computations of quasiparticle properties
of real materials.

In most of these studies the plane wave expansion of the LDA
pseudopotential basis-set is used, which makes an analysis in terms of
chemically relevant orbitals difficult.  An additional difficulty with
doing the GWA to the self-energy in a plane wave basis set
is that the computational effort  for systems with localized electrons
is enormous since the plane wave expansion of the wave function becomes 
a real problem especially for systems of localized electrons or 
low dimensional materials.

In this paper, we   propose a  new implementation of the GWA 
based on an all-electron method using the recently developed all-electron
projector
augmented wave (PAW) method.\cite{Blochl}
Our constructed  ab-initio method of quasiparticles is then based on the
framework of the LDA theory in conjunction with the PAW method.
The knowledge of the
one-electron Green function provided by the  PAW method allows us to
construct the quasiparticle self-energy within the GWA,
in which the dynamical screening of the electron-electron
interaction arises from a plasmon model  dielectric 
function\cite{Horsch,Hamada,Engel}  for which
the parameters are adjusted to the dielectric function calculated in the
random-phase-approximation \cite{Adler} (RPA).  The full
calculation of systems of many atoms in the unit cell is prohibitive on
sequential computers, so we had to parallelize our numerical code so that
the calculations become feasible on scalable parallel architectures.

Our paper is organized as follows: In the fist part we introduce the projector
augmented wave method which is used to solve the Kohn-Sham equations and 
provides the one-electron Green's function and the RPA dielectric function 
which are then used  to compute the self-energy. We  then describe
our GW implementation based on the PAW method and show
the difficulties,   traps,   encountered when implementing an all-electron
 GWA.  In particular, we mention the difficulty related to the decoupling of 
the core and valence electrons. 
We discuss also in details the use of symmetry to reduce the computational time
of the self-energy and the dielectric function, as well as  the parallelization
of our numerical code. In the third section 
 we  apply our method to determine the electronic structure of two
distinct semiconductor groups: some small and medium-band-gap  
semiconductors: Si, GaAs, AlAs, InP,
Mg$_2$Si and some large-band-gap semiconductors (insulator): 
C, and (LiCl). We then compare our results with available GWA
 calculations and experiments.

\section{Method of calculation}
\subsection{A brief introduction of the  PAW method}
In the density functional theory implemented in the framework of the LDA 
\cite{Hohenberg},
 an electronic structure calculation requires the solution of  Kohn-Sham 
type of equations in a self consistent way. 
The computation of material's band structure consists in finding
the Bloch wave functions $\Psi_{n{\bf k}}(r)$, where $n$ and ${\bf k}$ 
denote a band index and
a wave vector in the Brillouin zone, respectively. 
In the projector augmented plane wave (PAW) formalism, \cite{Blochl}
 all calculations
are performed on a smooth ``pseudo" wave function 
$\tilde{\Psi}_{n{\bf k}}(r)$ which is expressed
as a linear combination of plane waves. The passage from the smooth 
pseudo-wave function to the
all-electron wave function exhibiting the correct nodal behavior in the 
augmentation
regions (spheres centered on each atom) is achieved by defining three atomic 
type of functions in each
augmentation region:  (1)  The all-electron basis functions $\Phi_{i}^{a}(r)$,
(2)  The 'pseudo' basis functions $\tilde{\Phi}_{i}^{a}(r)$, 
(3) The projector functions $\tilde{p}_{i}^{a}(r)$.
Here $i=l_{i},m_{i},n_{i}$ where $l_{i}$ and $m_{i}$
 denote the orbital and magnetic quantum numbers, respectively.
 The index $n_{i}$ is introduced to leave the opportunity of choosing more than one function for
each angular momentum channel $(l_{i},m_{i})$.  These functions are 
defined so that:
\begin{equation}
\tilde{\Phi}_{i}^{a}(r)=\Phi_{i}^{a}(r) \qquad \textrm{for } r\geq r_{c}^{a} 
\end{equation}
where $r_{c}^{a}$ are  the radii of  non-overlapping spheres centered at each 
atomic site $a$.
The projector functions vanish for $ r\geq r_{c}^{a}$ and satisfy the 
orthogonality property:
\begin{equation}
\langle\tilde{p}_{i}^{a}|\tilde{\Phi}_{j}^{a}\rangle = \delta_{ij}
\end{equation}
Using these functions, the all-electron wave function $\Psi_{nk}(r)$ can be deduced
from the 'pseudo' wave function $\tilde{\Psi}_{nk}({\bf r})$ according to the relation
$\Psi_{n{\bf k}}({\bf r})=\tilde{\Psi}_{n{\bf k}}({\bf r})+\Psi_{n{\bf k}}^{1}({\bf r})
-\tilde{\Psi}_{n{\bf k}}^{1}({\bf r})$ with

\begin{equation}\label{decomposition}
\Psi_{n{\bf k}}^{1}({\bf r})-\tilde{\Psi}_{n{\bf k}}^{1}({\bf r})=
\sum_{a,i} [~ \Phi_{i}^{a}({\bf r-R^{a}})
-\tilde{\Phi}_{i}^{a}({\bf r-R^{a}})~] \langle \tilde{p}_{i}^{a}|
 \tilde{\Psi}_{n {\bf k}}\rangle
\end{equation}

where ${\bf R^{a}}$ denotes the atomic position of the atom $a$ 
in the unit cell.
It is useful to point out the fact that 
$\Psi_{n{\bf k}}^{1}-\tilde{\Psi}_{n{\bf k}}^{1}$
vanishes
in the interstitial region and defines the quantity necessary to describe the
 true wave function
in the augmentation regions while $\tilde{\Psi}_{n{\bf k}}$ describes 
the true wave function in the 
interstitial region.

\noindent
The PAW formalism is designed to easily calculate the expectation value 
of local or semi-local
 observables. For 
example the expectation value of an operator $A({\bf r})$ between 
two Bloch wave functions $\Psi_{n{\bf k}}$ 
and $\Psi_{m{\bf k}}$ can be calculated as a sum of three contributions:
\begin{equation}
A_{n{\bf k},m {\bf k}}=\langle\Psi_{n {\bf k}}|A|\Psi_{m {\bf k}}\rangle
=\tilde{A}_{n{\bf k},m{\bf k}}+A_{n{\bf k},m{\bf k}}^{1} -
\tilde{A}_{n{\bf k},m{\bf k}}^{1}
\end{equation}
where
\begin{equation}
\tilde{A}_{n{\bf k},m{\bf k}}=\langle\tilde{\Psi}_{n{\bf k}}|A|
\tilde{\Psi}_{m{\bf k}}\rangle
\end{equation}
This contribution is evaluated in the plane wave basis set. The last two 
contributions
\begin{equation}
A_{n{\bf k},m{\bf k}}^{1}=\sum_{i,j,a} \langle\tilde{\Psi}_{n{\bf k}}|
\tilde{p}_{i}^{a}\rangle
             \langle \Phi_{i}^{a}|A|\Phi_{j}^{a}\rangle 
             \langle \tilde{p}_{j}^{a}|\tilde{\Psi}_{m{\bf k}}\rangle
\end{equation}
and 
\begin{equation}
\tilde{A}_{n{\bf k},m{\bf k}}^{1}=\sum_{i,j,a} \langle\tilde{\Psi}_{n{\bf k}}|
\tilde{p}_{i}^{a}\rangle
             \langle \tilde{\Phi}_{i}^{a}|A|{\tilde\Phi}_{j}^{a}\rangle
             \langle \tilde{p}_{j}^{a}|\tilde{\Psi}_{m{\bf k}}\rangle
\end{equation}
are evaluated in the augmentation regions. 

\noindent
Bl\"ochl\cite{Blochl} managed to decompose the total valence energy into three 
contributions in complete
analogy with the decomposition of the expectation value of an operator
\begin{equation}
E={\tilde E} + E^{1} - {\tilde E}^{1}
\end{equation}
where $E$ includes the kinetic energy of the valence electrons, the 
interaction energy
with the nuclei having atomic number $Z^{a}$ and with the core electrons, 
which are described in the frozen core approximation, the Hartree energy and
 the exchange correlation energy of the valence electrons.

\noindent
Minimizing the total energy with respect to the $\tilde{\Psi}_{n{\bf k}}$
 (variationnal principle)
leads to a generalized eigenvalue problem which is solved in a
 self-consistent way, giving the
pseudo wave functions $\tilde{\Psi}_{n{\bf k}}$ from which the all-electron
 wave functions 
$\Psi_{n{\bf k}}$ are easily deduced by means of equation \ref{decomposition}.

\subsection{The GW approximation}
\subsubsection{Quasiparticle calculation in the GW approximation}
Solving the Kohn-Sham equations, where the exchange-correlation effects are 
included
in a mean way in an exchange-correlation potential $V_{xc}^{LDA}(r)$ 
yields to eigenvalues
which can not be assimilated to the excitation energies of the solids. 
Indeed, there isn't
any Koopman's theorem for the Kohn-Sham theory and these eigenvalues 
only have meaning
as Lagrange
parameters. The Green's function theory is a good tool for properly 
describing the
excitation energies. In the quasiparticle approximation, we can 
find the excitation
energies of the system by solving a quasiparticle equation
\begin{equation}
(T+V_{ext}+V_{h})\psi_{{\bf k}n}({\bf r}) + \int d^3r^{\prime}
 \Sigma({\bf r},{\bf r}^{\prime},E_n({\bf k}))\psi_{{\bf k}n}({\bf r}^{\prime}) =
 E_n({\bf k})\psi_{{\bf k}n}({\bf r})
\end{equation} 
instead of locating the poles of the Green's function. Here, T is the kinetic 
energy operator
($-\frac{1}{2}\nabla^{2}$ in atomic units), $V_{ext}$ is the external (ionic) 
potential, $V_{h}$
is the Hartree potential due to the average Coulomb repulsion of the electrons
 and $\Sigma$
is the self-energy operator which summarizes the many-body effects. $\Sigma$
 is in general
a non local, energy-dependent, non-Hermitian operator. The 
non-Hermitian part of $\Sigma$
gives rise to complex eigenvalues $E_n({\bf k})$. The real part of 
the eigenvalue
is associated with the energy of the quasiparticle while the imaginary part is 
related to the 
inverse of the lifetime of the quasiparticle. Fortunately, the imaginary 
part of the
eigenvalue, which can be attributed to the possibility for the quasiparticle 
to lower its
energy via particle-hole formations (the decay of the quasiparticle
 via emission
or absorption of plasmons is not effective near the Fermi level), becomes 
small near the Fermi level
 because of the restriction
of the phase space for these processes. This is known as Pauli blocking.
 So, the quasiparticle 
concept is effective near the Fermi Level.

The similitude between the quasiparticle equation and the familiar 
self-consistent field
equation in the Kohn-Sham formulation of the density-functional theory   
\begin{equation}
(T+V_{ext}+V_{h}+V_{xc})\Psi_{{\bf k}n}({\bf r}) = \epsilon_n({\bf k} ) 
\Psi_{{\bf k}n}({\bf r})
\end{equation}
is striking if we set $\Sigma^{LDA}=\delta(E)\delta({\bf r},
{\bf r}^{\prime}) V_{xc}({\bf r})$ 
in the quasiparticle
equation. So the Kohn-Sham equation can be regarded as an approximation
 to the quasiparticle 
equation where the self-energy is approximated by a local and 
energy-independent potential.

A more realistic, but relatively simple approximation to the self-energy, 
which takes into 
account both non-locality and dynamic correlations, known as the GW 
approximation (GWA),
was initiated by Hedin \cite{Hedin1,Hedin2}. This approximation was 
originally derived
from a many-body perturbation theory as the first term in the expansion 
of the self-energy
in the screened interaction W. Within this scheme, the self-energy 
\begin{equation} \label{self_energy}
\Sigma({\bf r},{\bf r}^{\prime},\omega)=\frac{i}{2\pi}\int d\omega' 
G({\bf r},{\bf r}^{\prime},\omega+\omega^{\prime})e^{i\delta\omega^{\prime}}
W({\bf r},{\bf r}^{\prime},\omega^{\prime})
\end{equation}
is approximated by a convolution with respect to the frequency variable of the 
Green's
function with the screened interaction along the real axis. 
Here $\delta$ is a positive
infinitesimal which ensures that only the poles of the Green's function
 associated 
to occupied states contribute to the integral, $G$ is the best available
 Green's function 
and $W$ is the best available screened Coulomb interaction. Accordingly, 
$G$ is taken to be
the Green's function built from LDA orbitals
\begin{equation}
G({\bf r},{\bf r}^{\prime},\omega)=\lim_{\delta \rightarrow 0^+}
 \sum_{n{\bf k}}
\frac{\Psi_{{\bf k}n}({\bf r})\Psi_{{\bf k}n}^*({\bf r}^{\prime})}
{\omega-\epsilon_n({\bf k})+i\delta sgn(\epsilon_n({\bf k})-\mu)}
\end{equation} 
where $\mu$ denotes the chemical potential, and where $ \delta \rightarrow
0^+$ means $\delta$ being a positive infinitesimal number.

The dynamically screened interaction is defined by
\begin{equation}\label{screened_interaction} 
W({\bf r},{\bf r}^{\prime},\omega)=\int d{\bf r}^{\prime\prime}
 \epsilon^{-1}({\bf r},{\bf r}^{\prime\prime},\omega) 
v({\bf r}^{\prime\prime},{\bf r}^{\prime})
\end{equation}
where $v$ denotes the bare Coulomb
interaction and $\epsilon$ the dielectric function defined by
\begin{equation}
\epsilon({\bf r},{\bf r}^{\prime},\omega)=\delta({\bf r}-{\bf r}^{\prime}) 
-\int v({\bf r},{\bf r}^{\prime\prime}) 
P^{0}({\bf r}^{\prime\prime},{\bf r}^{\prime},\omega) d{\bf r}^{\prime\prime}
\end{equation}
where the irreducible polarisability $P^{0}$ is calculated within the RPA
approximation (Bubble approximation)
\begin{equation}
P^{0}({\bf r},{\bf r}^{\prime},\omega)=-\frac{i}{2\pi}
\int d\omega^{\prime} G({\bf r},{\bf r}^{\prime},\omega)
 G({\bf r}^{\prime},{\bf r},\omega^{\prime}-\omega) e^{i\delta\omega^{\prime}}
\end{equation}
Using the LDA expression of the Green's function and integrating in the complex
plane leads to the familiar expression of the irreducible polarisability
\begin{equation} \label{irreducible_polarisability}
P^{0}({\bf r},{\bf r}^{\prime},\omega)=\sum_{l{\bf k},m{\bf k}^{\prime}}
\frac{[n_l({\bf k})-n_m({\bf k^{\prime}})]\Psi_{{\bf k}l}({\bf r})
\Psi_{{\bf k}l}^*({\bf r}^{\prime})\Psi_{{\bf k}^{\prime}m}({\bf r}^{\prime})
\Psi_{{\bf k}^{\prime}m}^*({\bf r}^{\prime})}
{\epsilon_l({\bf k})-\epsilon_m({\bf k}^{\prime})
-\omega+i\delta[n_l({\bf k})-n_m({\bf k}^{\prime})]}
\end{equation}
After having defined the different quantities, which enter the self-energy
 in the
GW approximation, it is natural to look for the solution of the quasiparticle
equation along the following line
\begin{equation}
\psi_{{\bf k}m}(r)=\sum_n\alpha_{mn}({\bf k})\Psi_{{\bf k}n}(r)
\end{equation}
It has been shown by inspection that, $\alpha_{mn} \simeq \delta_{mn}$, 
which  means that the quasiparticle
Hamiltonian is virtually diagonal in the $\Psi_{{\bf k}n}$ basis for
 semiconductors such
as Si\cite{Hybertsen1}. A plausible argument for  neglecting the non-diagonal 
part
has been given recently by Hedin.\cite{Hedin3} Since 
the quasiparticle wave functions 
$\psi_{{\bf k}m}$ are  similar 
to the LDA wave functions $\Psi_{{\bf k}n}$ (the overlap between the
quasiparticle wave function and the LDA wave function is about
99 \%\cite{Hybertsen1}), the numerical work is therefore considerably
reduced. Indeed, using first order perturbation theory yields the 
following result:
\begin{equation}\label{quasiparticle_energy}
E_n({\bf k})=\epsilon_n({\bf k})+\langle\Psi_{{\bf k}n}|
\Sigma({\bf r},{\bf r}^{\prime},E_n({\bf k}))|\Psi_{{\bf k}n}\rangle
- \langle\Psi_{{\bf k}n}|V_{xc}^{LDA}({\bf r})|\Psi_{{\bf k}n}\rangle
\end{equation}
In principle, the solution to this equation should be obtained via an
 iterative method, but
expanding Eq. (\ref{quasiparticle_energy}) to first order in energy 
around 
$\epsilon_n({\bf k})$ yields accurate numerical results. Then the 
quasiparticle energy can 
be obtained via
\begin{equation}\label{quasiparticle_energy_final}
E_n({\bf k})-\epsilon_n({\bf k})=Z_{n{\bf k}}[\langle\Psi_{{\bf k}n}|
\Sigma({\bf r},{\bf r}^{\prime},\epsilon_n({\bf k}))|\Psi_{{\bf k}n}\rangle
- \langle\Psi_{{\bf k}n}|V_{xc}^{LDA}(r)|\Psi_{{\bf k}n}\rangle]
\end{equation}
where the quasiparticle renormalization factor $Z_{n{\bf k}}$ is
\begin{equation}\label{Renormalization}
Z_{n{\bf k}}=[1-\langle\Psi_{{\bf k}n}|
\frac{\partial}{\partial\omega} \Sigma({\bf r},{\bf r}^{\prime},
\epsilon_n({\bf k}))
|\Psi_{{\bf k}n}\rangle]^{-1}
\end{equation}
Since $\frac{\partial}{\partial\omega} \Sigma < 0$, we have $0<Z_{n{\bf k}}<1$ 
with typical
values of 0.8 for bands close to the band gap and for all materials 
considered here (see Table \ref{renormalization}). Values
of $Z_{n{\bf k}}$ about 0.8 imply that we still have well-defined 
quasiparticles in the system
but that 20 percents of the spectral weight is now distributed over a range
 of frequencies.

\subsubsection{Calculating the matrix elements of $\Sigma$ within the PAW
 formalism}
As can be seen from Eq. (\ref{quasiparticle_energy_final}), the central 
problem of
this scheme consists in evaluating the diagonal matrix elements of the 
self-energy between
LDA orbitals.
The quantities which enter the self-energy are functions $f$ of two 
locations ${\bf r}$ and
${\bf r}^{\prime}$. These functions have the following translational 
symmetry property 
$f({\bf r}+{\bf R},{\bf r}^{\prime}+{\bf R})= f({\bf r},{\bf r}^{\prime})$ 
where $R$ is
a Bravais lattice vector. Then, we can fix the Fourier transform convention 
for such functions
\begin{equation}
f({\bf r},{\bf r}^{\prime}, \omega)=
\frac{1}{\Omega}\sum_{{\bf q},{\bf G},{\bf G}^{\prime}}
e^{i({\bf q}+{\bf G}).r}f_{{\bf G}{\bf G}^{\prime}}({\bf q},\omega)
e^{-i({\bf q}+{\bf G}^{\prime}).{\bf r}^{\prime}}
\end{equation}
where ${\bf q}$ is a wave vector in the first Brillouin zone, ${\bf G}$
 a reciprocal
lattice vector and $\Omega$ the crystal volume.

In the plane wave basis, the dielectric function can be defined as
\begin{equation} \label{epsilon}
\epsilon_{{\bf G}{\bf G}^{\prime}}({\bf q},\omega)=
\delta_{{\bf G}{\bf G}^{\prime}}
 - v({\bf q}+{\bf G})P^{0}_{{\bf G}{\bf G}^{\prime}}({\bf q},\omega)
\end{equation}
Then according to Eqs. (\ref{self_energy} and \ref{screened_interaction}), 
the dielectric
matrices $\epsilon_{{\bf G}{\bf G}^{\prime}}({\bf q},\omega)$ have 
to be calculated 
and inverted for many values of $\omega$. This is computationally time 
consuming. 
Nevertheless, it has been carried out by some
authors\cite{Godby1,arya} who choose to evaluate the frequency 
integral by using a Gaussian integration scheme along
the imaginary axis to circumvent the problem of the pole structure of
 the screened interaction along 
the real frequency axis\cite{Godby1}. An alternative approach is the use of a
 plasmon pole model\cite{Hybertsen1,Horsch,Hamada,Engel}
 to mimic the frequency dependence of the dielectric matrix. 
These models give a good description
of the low frequency behavior of the dynamically screened interaction and 
allow the determination of  an
analytic expression for  the frequency integral appearing in Eq. 
(\ref{self_energy}). We used 
three types of plasmon pole models to describe approximately the dependence
of $\epsilon^{-1}(\omega)$ on the frequency $\omega$. These different models 
were proposed by   Von der Linden and Horsch \cite{Horsch},
 Engel and Farid\cite{Engel}, and  by Hamada {\it et al.}\cite{Hamada}
 We choose to detail the first one  and emphasize that the values of the 
quasiparticle 
energies are not  sensitive to the model type (see Table
\ref{qp_si_3_plasmon} which details the calculated QP energies of Si with
these three models).

In such an approach, we consider the symmetrized dielectric matrix 
$\tilde{\epsilon}_{{\bf G}{\bf G}^{\prime}}({\bf q},\omega)$ defined by
\begin{equation} \label{symetrised_epsilon}
\tilde{\epsilon}_{{\bf G}{\bf G}^{\prime}}({\bf q},\omega) = 
\frac{|{\bf q}+ {\bf G}|}{|{\bf q}+{\bf G}^{\prime}|}
{\epsilon}_{{\bf G}{\bf G}^{\prime}}({\bf q},\omega)
\end{equation}
Introducing the following notation
\begin{equation} \label{matrix_element}
M^{nm}_{{\bf G}}({\bf k},{\bf q})=
\langle\Psi_{{\bf k}-{\bf q}n}|e^{-i({\bf q}+{\bf G}).{\bf r}}|
\Psi_{{\bf k}m}\rangle
\end{equation}
and using Eqs. (\ref{symetrised_epsilon}, \ref{epsilon}, 
\ref{irreducible_polarisability}) leads
the following result for the static symmetrized dielectric matrix
\begin{equation}\label{full_symetrised_epsilon}
\tilde{\epsilon}_{{\bf G}{\bf G}^{\prime}}({\bf q},\omega=0) = 
\delta_{{\bf G}{\bf G}^{\prime}}
-\frac{16\pi}{\Omega|{\bf q}+{\bf G}||{\bf q}+{\bf G}^{\prime}|}
\sum_{v,c,{\bf k}} \frac{ M^{vc}_{{\bf G}}({\bf k},{\bf q}) 
[M^{vc}_{{\bf G}^{\prime}}({\bf k},{\bf q})]^{*}}
{\epsilon_{v}({\bf k}-{\bf q})-\epsilon_{c}({\bf k})}
\end{equation}
We then diagonalize the hermitian $\tilde{\epsilon}({\bf q},\omega=0)$. 
Its real eigenvalues $\lambda_p({\bf q})$
and orthonormal
eigenvectors $|\phi_p({\bf q})\rangle$ can be used to perform the matrix 
inversion according to
\begin{equation}
\tilde{\epsilon}^{-1}({\bf q},\omega=0)=\sum_p |\phi_p({\bf q})\rangle
 \frac{1}{\lambda_p({\bf q})}
\langle \phi_p({\bf q})|
\end{equation}
Von der Linden and Horsch 
supposed that the frequency dependency of the dielectric matrix is 
encompassed in  the eigenvalues
of the static symmetrized dielectric matrix while the eigenvectors are 
frequency independent
\begin{equation}
\tilde{\epsilon}^{-1}_{{\bf G}{\bf G}^{\prime}}({\bf q},\omega)=
\sum_p \phi_{\bf G}^p(q) \frac{1}{\lambda_p({\bf q},\omega)}
[\phi_{{\bf G}^{\prime}}^p({\bf q})]^*
\end{equation}
where
\begin{equation} \label{inverse_eigenvalue}
\frac{1}{\lambda_p({\bf q},\omega)}=1+\frac{z_p({\bf q})}{2}\omega_p({\bf q})
\{\frac{1}{\omega-(\omega_p({\bf q})-i\delta)}-
\frac{1}{\omega+(\omega_p({\bf q})-i\delta)}\}
\end{equation}
It should be noted that such a parameterization of the frequency 
dependence 
of the inverse
of the eigenvalues guarantees that $\tilde{\epsilon}$ is an even 
function of $\omega$. Here,
two parameters have to be determined:  (1) The force of the pole 
$z_p({\bf q})$ which is defined
by letting $\omega=0$ in Eq. (\ref{inverse_eigenvalue})
\begin{equation} \label{force}
z_p({\bf q})=1-\lambda_p^{-1}({\bf q},0),
\end{equation}
and is  positive since it can be shown that $\lambda_p^{-1}({\bf q},0)$ lies
in the interval $(0,1)$\cite{Car}. 
(2) The frequency of the pole $\omega_p({\bf q})$ 
which is determined from the Johnson's sum rule\cite{Johnson}.
If we introduce the following quantities
\begin{equation} \label{theta_def}
\Theta_{\bf G}^p({\bf q})=\frac{\phi_{{\bf G}}^p({\bf q})}{|{\bf q}+{\bf G}|},
\end{equation}
and 
\begin{equation}
L_{{\bf G}{\bf G}^{\prime}}=({\bf q}+{\bf G})({\bf q}+{\bf G}^{\prime})
\rho({\bf G}-{\bf G}^{\prime}),
\end{equation}
where $\rho({\bf G})$ denotes a Fourier component of the charge density,
we can write the frequencies of the poles according to
\begin{equation} \label{frequency}
\omega_p({\bf q})^2=\frac{4\pi}{z_p({\bf q})}
\langle\Theta^p({\bf q})| L({\bf q}) |\Theta^p({\bf q})\rangle
\end{equation}
Using the usual development of $\rho$ and $\Theta$ along the reciprocal vectors,
it can be easily shown that
\begin{equation} \label{theta}
\langle\Theta^p({\bf q})| L({\bf q}) |\Theta^p({\bf q})\rangle=
\frac{1}{V}\int\rho({\bf r})|\nabla\Theta({\bf r})|^2d^3{\bf r}
\end{equation}
where $V$ denote the volume of the unit cell. Since $z_p({\bf q})$ and the
 quantity defined by Eq. (\ref{theta})
are both positive we deduce that the frequency defined by Eq. (\ref{frequency})
is positive. Now all the ingredients of the plasmon pole model have been
 defined, and we are
able to give an analytic expression of the matrix elements of the self-energy 
in the GW approximation between LDA Bloch orbitals.

\begin{figure}
\begin{minipage}{6.0in}
\epsfysize=7.0in
\centerline{\epsfbox{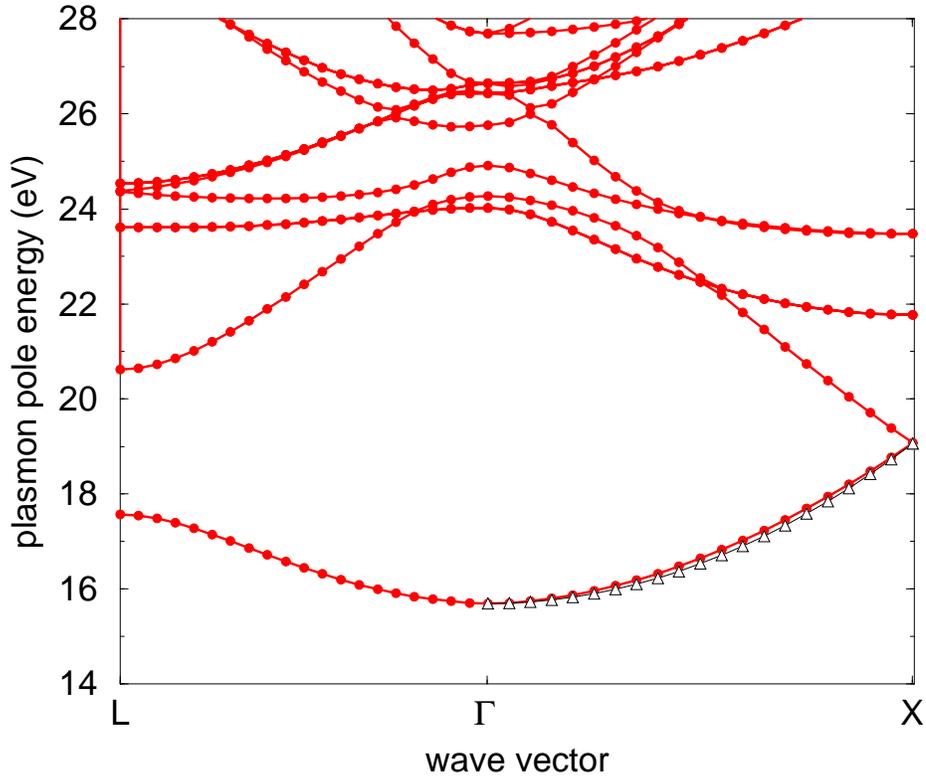}}
\end{minipage}
\caption{
\label{engel_farid}
Calculated Engel-Farid Plasmon model band structure  of Si along L, 
$\Gamma$, and X high symmetry directions.
The agreement with the results of Engel and Farid \protect{\cite{Engel}}, 
and Aulbur \protect{\cite{aulbur_thesis}} are excellent. For small {\bf k} wave
vectors the lowest plasmon band shows a quadratic dispersion (up-triangle
curve) $\omega_{0}({\bf k})=\omega_{0}({\bf 0})+\alpha |{\bf k}|^{2}$,
with a dimensionless
direction-dependent dispersion coefficient $\alpha$.
We find $\omega_{0}({\bf 0})=$ 15.7 eV and $\alpha_{\Delta}=$ 0.33 in good
agreement with the values of 15.91 eV and 0.34 of Engel 
and Farid\protect{\cite{Engel}} as well as the
experimental values \cite{raether}  of 16.7 eV and 0.41.  }
\end{figure}

\begin{figure}
\begin{minipage}{6.0in}
\epsfysize=7.0in
\centerline{\epsfbox{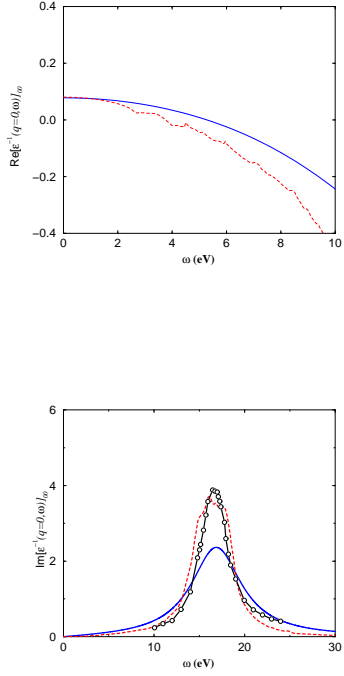}}
\end{minipage}
\caption{
\label{hamada_model}
Ab-initio calculated  real and imaginary parts of the inverse dielectric 
function (dashed curve)  of Si compared to the plasmon pole model of
Hamada {\it et al} (solid line) and with experiment (open circles)
\protect{\cite{raether}}.
}
\end{figure}

To test our implementation of the plasmon pole model, we have plotted in Fig.
\ref{engel_farid} the Engel-Farid Plasmon model band structure  of Si along
L, $\Gamma$, and X high symmetry directions. We have found that 
our implementation is in excellent agreement with the results
of Engel and Farid \protect{\cite{Engel}}, and Aulbur 
\cite{aulbur_thesis}. Indeed, for small ${\bf k}$ wave vectors, the lowest 
plasmon band
shows a quadratic dispersion 
$\omega_{0}({\bf k})=\omega_{0}({\bf 0})+\alpha |{\bf k}|^{2}$, with a dimensionless 
direction-dependent dispersion coefficient $\alpha$ (see up-triangle
curve in Fig. \ref{engel_farid}). 
We find $\omega_{0}({\bf 0})=$15.7 eV and $\alpha_{\Delta}=$0.33 in good agreement
with the values of 15.91 eV and 0.34 of Engel and Farid as well as the experimental
values of 16.7 eV and 0.41\cite{raether}.
To check also the validity of the plasmon pole model
to be used as a substitute for the dielectric function of real materials 
we have compared the plasmon pole model of Hamada et al.\cite{Hamada} with
our direct ab-initio computation of the dielectric function  
 within the RPA including the so called 
local-field effects (see Fig. \ref{hamada_model}) and with available 
experimental results\cite{raether} for the energy loss spectrum. We notice 
that the model mimics nicely our ab initio calculated dielectric function.   

The matrix elements of the self-energy could be divided into an energy independent
contribution $\Sigma^{hf}$ and an energy dependent contribution $\Sigma(\omega)$. The
first contribution corresponds to the Hartree-Fock contribution and is given by
\begin{equation}\label{hf_sigma}
\langle \Psi_{{\bf k}n}|\Sigma^{hf}|\Psi_{{\bf k}n}\rangle=
-\frac{4\pi}{\Omega}\sum_{\bf q}\sum_{m~ occ}\sum_{{\bf G}}
\frac{ |M_{{\bf G}}^{mn}({\bf k},{\bf q})|^{2}} {|{\bf q}+{\bf G}|^2}
\end{equation} 
where the summation runs only over the occupied states. The
energy dependent contribution can be expressed as
\begin{equation} \label{correlation_sigma}
\langle\Psi_{{\bf k}n}|\Sigma(\omega)|\Psi_{{\bf k}n}\rangle 
=\frac{4\pi}{\Omega} \sum_{{\bf q}, m, p}  
\frac{z_{p}({\bf q})\omega_{p}({\bf q})/2}
{\omega-\epsilon_{m}({\bf k}-{\bf q})+[\omega_{p}({\bf q})-i\delta] 
sgn(\mu-\epsilon_{m}({\bf k}-{\bf q}))} |\beta^{mn}_{p}({\bf k},{\bf q})|^{2}
\end{equation}
where
\begin{equation}
\beta^{mn}_p({\bf k},{\bf q})=\sum_{{\bf G}}[M^{mn}_{\bf G}({\bf k},{\bf q})]^*
\Theta_{p{\bf q}}({\bf G})
\end{equation}
and
\begin{displaymath}
sgn(\mu-\epsilon_m({\bf k}-{\bf q}))=\left\{ \begin{array}{ll}
1 & \textrm{for~}\epsilon_m({\bf k}-{\bf q})<\mu\\
-1 & \textrm{for~}\epsilon_m({\bf k}-{\bf q})>\mu\\
 \end{array} \right.
\end{displaymath}
Here $m$ denotes an electronic band index, $p$ a plasmon band index and ${\bf q}$ 
a vector in the
Brillouin zone. It should be emphasized that the summation over $m$ is not 
restricted 
to occupied states as in the expression of the
Hartree-Fock contribution. Both the poles of the Green's function and of the 
screened interaction
contribute to this expression.
\subsubsection{Numerical details}
One of the central problem within the realization of the GW approximation
is the calculation of matrix elements whose type is defined by Eq. (\ref{matrix_element}).
Using the PAW formalism, the smooth 'pseudo wave function' $\tilde{\Psi}_{{\bf k} n}$
 associated to an 'all-electron' LDA wave function $\Psi_{{\bf k} n}$ can be written
\begin{equation}
\tilde{\Psi}_{{\bf k} n}=\frac{1}{\sqrt{V}} 
\sum_{{\bf G}} A_{{\bf k} n}({\bf G}) e^{i({\bf k}+{\bf G}).{\bf r}}
\end{equation}
where the sum runs over reciprocal lattice vectors. As illustrated in the part dedicated
to the PAW formalism, 
the expectation value of $e^{-i({\bf q}+{\bf G}).{\bf r}}$ can be
divided into three parts according to $M^{nm}_{{\bf G}}({\bf k},{\bf q})=
\langle \tilde{\Psi}_{{\bf k}-{\bf q} n}|e^{-i({\bf q}+{\bf G}).{\bf r}}
|\tilde{\Psi}_{{\bf k} m}\rangle + 
\langle \Psi_{{\bf k}-{\bf q} n}^{1}|e^{-i({\bf q}+{\bf G}).{\bf r}}
|\Psi_{{\bf k} m}^{1}\rangle
-\langle \tilde{\Psi}_{{\bf k}-{\bf q} n}^{1}|e^{-i({\bf q}+{\bf G}).{\bf r}}
|\tilde{\Psi}_{{\bf k} m}^{1}\rangle $. 
The first term, which involves plane waves is defined 
as
\begin{equation}
\sum_{{\bf G}^{\prime}}A_{{\bf k}-{\bf q} n}^{*}({\bf G}^{\prime}) A_{{\bf k} m}
({\bf G}+{\bf G}^{\prime})  
\end{equation}
In our scheme, the summation over reciprocal lattice vectors in this expression 
takes account
of a vector ${\bf G}^{\prime}$ if both 
${\bf G}^{\prime}$ and ${\bf G}+{\bf G}^{\prime}$
are smaller than a cutoff parameter. In general 300 ${\bf G}^{\prime}$
 vectors were
included in the summation for the systems studied here, except for Diamond 
where the convergence is achieved only when 400 ${\bf G}^{\prime}$ were used.
The two remaining terms which involve localized contributions 
can be expressed as
\begin{equation}
\sum_{{\bf a}, i, j} 
\langle\tilde{\Psi}_{{\bf k}-{\bf q} n}|\tilde{p}_{j}^{\bf a}\rangle
[~\langle \Phi_{j}^{\bf a}|e^{-i({\bf q}+{\bf G}).{\bf r}}|
\Phi_{i}^{\bf a}\rangle
-\langle \tilde{\Phi}_{j}^{\bf a}|e^{-i({\bf q}+{\bf G}).{\bf r}}|
{\tilde\Phi}_{i}^{\bf a}\rangle~]
\langle \tilde{p}_{i}^{\bf a}|\tilde{\Psi}_{m{\bf k}}\rangle
\end{equation}
Since the overlap between the 'pseudo wave functions' and the projectors is known, we
have to calculate quantities like $\langle \Phi_{j}^{\bf a}|e^{-i({\bf q}+{\bf G}).{\bf r}}|
\Phi_{i}^{\bf a}\rangle$. Using the development of plane waves along 
spherical Bessel functions $j_{l}$, we get
\begin{equation}
\langle \Phi_{j}^{\bf a}|e^{-i({\bf q}+{\bf G}).{\bf r}}|\Phi_{i}^{\bf a}\rangle = 
\left\{ \begin{array}{ll}
4\pi~ e^{-i({\bf q}+{\bf G}).\bf{R}_{{\bf a}}} 
~\sum_{lm}~ (-i)^{l}~ Y_{lm}(\widehat{{\bf q}+{\bf G}})
~G^{lm}_{l_{i}m_{i}~l_{j}m_{j}}  \\
\times \int_{0}^{r_{c}^{\bf a}} d{\bf r}~ r^{2} 
~j_{l}(|{\bf q}+{\bf G}||{\bf r}|)~ \Phi_{l_{j}n_{j}}(r)~ \Phi_{l_{i}n_{i}}(r)  \\
\end{array}\right.
\end{equation}
where the Gaunt coefficients are defined to be
\begin{equation}
G^{lm}_{l_{i}m_{i}~l_{j}m_{j}} = 
\sqrt{4 \pi} \int d\Omega~ Y_{l_{i}m_{i}}^{*}(\widehat{{\bf r}})
~Y_{lm}^{*}(\widehat{{\bf r}})~ Y_{l_{j}m_{j}} (\widehat{{\bf r}})
\end{equation}

The symmetrized dielectric matrix defined by 
Eq.(\ref{full_symetrised_epsilon}) as well
 as the matrix elements of the
self-energy defined by Eqs. (\ref{hf_sigma}) and (\ref{correlation_sigma})
are obtained by an integration over the Brillouin zone. We used
the special points' technique\cite{Monkhorst} in which the summation over a uniform
mesh of ${\bf k}$-points is reduced by symmetry to a summation over fewer 
special ${\bf k}$-points if
the integrand possesses the full symmetry of the point group of the lattice. The summation
over ${\bf q}$-points in the expectation value of the self-energy has to
 be carried out  carefully, 
since the integrands have an integrable singularity in $\frac{1}{{\bf q}^{2}}$ 
for ${\bf q} \rightarrow 0$. This can be readily seen in the expectation
value of $\Sigma^{hf}$ defined by Eq. (\ref{hf_sigma}) where the divergence occurs
when ${\bf G}=0$. We follow the idea of Gygi and Baldereschi\cite{Gygi} introduced for
fcc lattice by considering a smooth
function $F({\bf q})$ which reflects the translational symmetry of the Bravais lattice and 
which diverges as $\frac{1}{{\bf q}^{2}}$ as $q$ vanishes. If we have to integrate a function
$g({\bf q})$ which behaves as $\frac{A}{{\bf q}^{2}}$ for small ${\bf q}$, we can write
\begin{equation}
\sum_{{\bf q}}g({\bf q})=\sum_{{\bf q}} \left[g({\bf q})-A~F({\bf q})\right] 
+ A \sum_{{\bf q}} F({\bf q})
\end{equation}
Such a decomposition is illustrated for the Hartree-Fock contribution
\begin{equation}
\langle \Psi_{{\bf k}n}|\Sigma^{hf}|\Psi_{{\bf k}n}\rangle=
-\frac{4\pi}{\Omega}\sum_q\sum_{m~ occ}\left[\sum_{{\bf G}}
\frac{ |M_{{\bf G}}^{mn}({\bf k},{\bf q})|^{2}} {|{\bf q}
+{\bf G}|^2}-F({\bf q})\delta_{mn}\right] 
-\frac{4\pi}{\Omega} \sum_{m~ occ} \delta_{mn} \sum_q F({\bf q})
\end{equation}
The  function in square brackets does not contain any divergence owing to 
the following property
\begin{equation}
\lim_{{\bf q} \rightarrow 0} M_{{\bf 0}}^{mn}({\bf k},{\bf q})=\delta_{mn}
\end{equation}
and is easily integrated by special points method while the integral of $F({\bf q})$ over the 
Brillouin zone is performed analytically. The same type of method is used to treat the 
$\frac{1}{{\bf q}^{2}}$ singularity in Eq. (\ref{correlation_sigma}). Here the development
of $|\beta^{mn}_{p}({\bf k},{\bf q})|^{2}$ shows that a divergence like $\frac{1}{q}$ also
occurs. The problem can still be solved by using another function which diverges as $\frac{1}{q}$
if ${\bf q} \rightarrow 0$. As a matter of fact, this divergence is less severe than the previous
one and does not require a special treatment since the accuracy of the numerical results is
not affected if we neglect this divergence. It should be noted that the treatment of the
singularity in Eq. (\ref{correlation_sigma}) necessitates the evaluation
of the symmetrized dielectric
matrix for ${\bf q} \rightarrow 0$. As the convergence of the head element of this matrix as a 
function of the number of ${\bf k}$-points is slow, the calculation is performed separately. All
other Brillouin zone integrations are carried out using 10 special 
${\bf k}$-points, which produces accurate numerical results (see the result section).

The evaluation of the quasiparticle energies requires the  determination
of  the renormalization
factor defined by Eq. (\ref{Renormalization}). The derivative of the self-energy is then
evaluated by using a finite difference scheme with a step equal to 1 eV.
The values of
the renormalization factors are summarized in Table \ref{renormalization} for the different
materials studied here. These constants $Z$ are roughly the same for the electron and hole
states. Furthermore $Z$ is similar for all the materials considered here. The values indicate
that account of dynamical renormalization is crucial to get quantitatively correct results.
At the same time values of $Z$ close to unity show that the quasiparticles are well defined
and that the GW approximation is reasonable.

The last point to be discussed is the evaluation of the matrix elements of the
exchange-correlation potential which appears in Eq. (\ref{quasiparticle_energy}). Here
some problem arise due to the non-linearity of the exchange-correlation potential. Indeed,
in the PAW calculation, the exchange correlation potential is calculated by taking into
account the valence $n_{v}$ and core electron density $n_{c}$ since 
$V_{xc}^{LDA}({\bf r})=V_{xc}^{LDA}[n_{v}({\bf r})+n_{c}({\bf r})]$. Therefore, the
self-energy in the GW approximation describe only the valence electrons, 
and we are
obliged to make the assumption that the core-valence exchange and core-polarization 
contributions to the energy of a valence state is given by\cite{Hybertsen1}
\begin{equation} \label{core_exchange}
\langle\Psi_{{\bf k}n}|V_{xc~core-val}|\Psi_{{\bf k}n}\rangle=
\langle\Psi_{{\bf k}n}|V_{xc}[n_{v}+n_{c}]|\Psi_{{\bf k}n}\rangle
-\langle\Psi_{{\bf k}n}|V_{xc}[n_{v}]|\Psi_{{\bf k}n}\rangle
\end{equation}   
Such a procedure is some what hidden when pseudo-potential are used since
such an operation is performed in the unscreening of the pseudo-potential
by subtracting $V_{xc}[n_{v}]$. The shortcoming of this approach is that
the ionic pseudo-potential is dependent on the valence configuration, reducing
the transferability of the potential. Moreover, it has been shown that including
core corrections  to the exchange and correlation is necessary to describe 
properly the structural properties of
solids\cite{LouieFroyen}. The PAW does not suffer this shortcoming but it seems
that justifying Eq. (\ref{core_exchange}) is not an easy task. Considering
that the argument mentioned before is correct, the quantity which must be subtracted
is then defined by
\begin{equation}
\langle\Psi_{{\bf k}n}|V_{xc}^{LDA}[n_{v}({\bf r})]|\Psi_{{\bf k}n}\rangle=
\left\{ \begin{array}{ll}
\langle\tilde{\Psi}_{{\bf k}n}|V_{xc}^{LDA}[{\tilde{n}}_{v}({\bf r})]
|\tilde{\Psi}_{{\bf k}n}\rangle 
+\sum_{i,j,a} \langle\tilde{\Psi}_{n{\bf k}}|\tilde{p}_{i}^{a}\rangle \\
\times \left [\langle \Phi_{i}^{a}|V_{xc}^{LDA}[n_{v}^{1}({\bf r})]|\Phi_{j}^{a}\rangle
-\langle {\tilde{\Phi}}_{i}^{a}|V_{xc}^{LDA}[{\tilde{n}}_{v}^{1}({\bf r})]
|{\tilde{\Phi}}_{j}^{a} \rangle\right ]
\langle \tilde{p}_{j}^{a}|\tilde{\Psi}_{n{\bf k}}\rangle \\
\end{array}\right.
\end{equation}
Table \ref{exchange} shows the matrix elements of the exchange-correlation
potential of Si calculated using the PAW formalism compared to the results obtained
by means of the LAPW method\cite{Hamada}. The agreement between our results and the
results of Hamada {\it et al.} is remarkably good if we 
consider that these results are
based on different calculation schemes and that different parameterizations of the
exchange-correlation energy are used.

\subsubsection{Use of symmetry to reduce computational cost}

If $\Psi_{{\bf k} n}$ is a Bloch wave function solution of the Kohn-Sham
equation and $R$ a symmetry operation belonging to the point group of the 
crystal
which we suppose to be symmorphic, 
we can write the way the Bloch wave function
transform under such a symmetry operation
\begin{equation}\label{matrix_element_property}
\Psi_{{\bf k} n}(R^{-1}{\bf r})=\sum_{m} D(R)_{nm} \Psi_{{\bf k} m}({\bf r}),
\end{equation} 
where $D(R)_{nm}$ denote the unitary transformation associated to the symmetry
operation $R$. If the state $\Psi_{{\bf k} n}$ is non degenerated, the transformation 
rule of the wave function simplify greatly since $D(R)_{nm}=\delta_{nm}$. We suppose
now that the states considered here are non degenerated to simplify the discussion. Using
such a relation it can be  shown that the matrix elements defined by 
Eq. (\ref{matrix_element})
satisfy the following relation
\begin{equation}
M^{nm}_{{\bf G}}({\bf k}, R{\bf q})=M^{nm}_{R^{-1}{\bf G}}({\bf k}, {\bf q})
\qquad{\textrm {for}}~ R{\bf k}={\bf k},
\end{equation} 
that means for the symmetry operations belonging to the little group $G_{\bf k}$ of
the point group $G$. Then, it can be proved that the integrand appearing in the
Hartree-Fock contribution defined by Eq. (\ref{hf_sigma}) is invariant under symmetry
operations belonging to  $G_{\bf k}$. Such a symmetry property reduces the 
number
of {\bf q}-points for which the integrand has to be calculated. 
Indeed, if $BZ_{{\bf k}}$ denotes
the irreducible Brillouin zone defined by $G_{\bf k}$, the Hartree-Fock contribution
can be rewritten as
\begin{equation}
\langle \Psi_{{\bf k}n}|\Sigma^{hf}|\Psi_{{\bf k}n}\rangle=
-\frac{4\pi}{\Omega}\sum_{{\bf q}\in BZ_{{\bf k}}} w({\bf q})\sum_{m~ occ}\sum_{{\bf G}}
\frac{ |M_{{\bf G}}^{mn}({\bf k},{\bf q})|^{2}} {|{\bf q}+{\bf G}|^2}
\end{equation}
where $w({\bf q})$ denotes the weight of the ${\bf q}$-point. It should be noted here that
the term we have subtracted from this expression to cancel the Coulomb singularity does 
not give rise to complications since this term is invariant under all 
symmetry operations  of $G$.
Moreover, if the state $\Psi_{{\bf k} n}$ is degenerated, we should sum the matrix elements
over all degenerated states to get a true invariant integrand. 
The same type of symmetry reduction holds for the calculation of the symmetrized static 
dielectric matrix using the fact that
\begin{equation}
M^{nm}_{{\bf G}}(R{\bf k}, {\bf q})=M^{nm}_{R^{-1}{\bf G}}({\bf k}, {\bf q})
\qquad{\textrm {for}}~ R{\bf q}={\bf q}
\end{equation}
and
\begin{equation}
\epsilon_{n}(R{\bf k})=\epsilon_{n}({\bf k})
\end{equation}
We then get
\begin{equation}
\tilde{\epsilon}_{{\bf G}{\bf G}^{\prime}}({\bf q},\omega=0) = \delta_{{\bf G}{\bf G}^{\prime}}
-\frac{16\pi}{\Omega|{\bf q}+{\bf G}||{\bf q}+{\bf G}^{\prime}|}
\sum_{{\bf k}\in BZ_{{\bf q}}}\sum_{v,c} \sum_{R \in G_{\bf q}}
\frac{ M^{vc}_{R{\bf G}}({\bf k},{\bf q})
[M^{vc}_{R{\bf G}^{\prime}}({\bf k},{\bf q})]^{*}}
{\epsilon_{v}({\bf k}-{\bf q})-\epsilon_{c}({\bf k})}
\end{equation}
The relationship between the matrix elements of $\tilde{\epsilon}_{{\bf G}{\bf G}^{\prime}}$
also reduces the computational cost. 
Using the symmetry property of the symmetrized dielectric function 
$\tilde{\epsilon}({\bf r},{\bf r}^{\prime})=\tilde{\epsilon}(R{\bf r},R{\bf r}^{\prime})$, 
it can be shown that
\begin{equation}
\tilde{\epsilon}_{{\bf G}{\bf G}^{\prime}}(R{\bf q},\omega=0)
=\tilde{\epsilon}_{R^{-1}{\bf G}R^{-1}{\bf G}^{\prime}}({\bf q},\omega=0).
\end{equation}
So both the hermiticity of $\tilde{\epsilon}$ and the relationship between the matrix elements 
which results from the symmetry operations leaving ${\bf q}$ invariant are 
used to reduce the number
of matrix elements to be computed. Now, we have to remember that the plasmon pole parameters are
obtained by solving an eigenvalue problem 
\begin{equation}
\int d{\bf r}^{\prime} ~\tilde{\epsilon}({\bf r},{\bf r}^{\prime}) 
~\phi^{p}({\bf q},{\bf r}^{\prime})
=\lambda_{p}({\bf q}) ~\phi^{p}({\bf q},{\bf r})
\end{equation} 
By analogy with the resolution of the Schr\"odinger type equation in a crystal, it can be shown
that
\begin{equation}
\phi_{{\bf G}}^{p}(R{\bf q})=\phi_{R^{-1}{\bf G}}^p({\bf q}) \qquad
{\textrm{and}} \qquad \lambda_p(R{\bf q})=\lambda_p({\bf q})
\end{equation}
These symmetry properties with Eqs. 
(\ref{force}, \ref{theta_def}, and \ref{frequency}) can be used
to show that 
\begin{equation}
z_{p}(R{\bf q})=z_{p}({\bf q}) \qquad
{\textrm{and}} \qquad \omega_p(R{\bf q})=\omega_p({\bf q})
\end{equation}
If the point group $G$ of the crystal does not contain the inversion symmetry, the
time reversal symmetry could also be implemented. Because of these symmetry relations,
the eigenvectors and eigenvalues of the symmetrized dielectric matrix are only computed 
for irreducible ${\bf q}$-points with respect to the point group of the crystal. Now
it can be demonstrated that the integrand appearing in the dependent energy contribution
to the matrix element of the self-energy defined by Eq. (\ref{correlation_sigma}) is
invariant under symmetry operations belonging to the little group of ${\bf k}$
denoted $G_{\bf k}$ as in the case of the Hartree-Fock contribution.

\subsubsection{Parallelization}
Because GWA is computationally involved the calculation of QP properties
in a sequential computer is time consuming. We have used the message passing
interface (MPI) to parallelize our numerical code on an IBM SP2. 
The most
straitforward  parallelization which we were  able to perform 
was any loop or a summation  involving  
{\bf k}-point mesh over the  Brillouin zone. First, the 
irreducible {\bf k}-point with respect to the point group of the 
crystal were distributed on different processors. Then each processor 
diagonalize the Hamiltonian for a certain number of {\bf k}-points. It
then evaluates the symmetrized dielectric function and calculates the
ingredients of the plasmon-pole model. Second, each processor calculates
the self-energy correction for a certain number of {\bf k}-points. and 
the results are gathered by the root processor to get the GW correction. 
This simple parallelization scheme made our computer code run much fast on the 
IBM SP2. Our choice of the the {\bf k}-point mesh was also motivated by a 
future parallelization of the LDA eigenvalue problem for systems with many
atoms per unit cell. We can set different pools of processors, and each
pool will produce the  diagonalization of a large Hamiltonian in parallel 
making the code highly scalable with the number of atoms of the system. 

\section{GW quasiparticle results}
\subsection{Quasiparticle results for small and medium band-gap semiconductors: 
Si, GaAs, AlAs, InP, and Mg$_2$Si} 

In this subsection we present the electronic structure of several
small and medium gap semiconductors which are used as tests for the 
implementation of our  all-electron PAW-GWA method. 
As mentioned earlier, 
we have implemented three different types of plasmon-pole models available
in the literature, \cite{Horsch,Hamada,Engel} and compared the 
resulting quasiparticle energies of Si
for high symmetry points. Table \ref{qp_si_3_plasmon} 
shows that these quasiparticle energies
are not sensitive to the type of plasmon-pole model used to mimic the
frequency dependence of the screened interaction. This trend has been
confirmed for the other semiconductors studied in this paper and was
already mentioned in the seminal work of Hedin \cite{Hedin2} on the
Jellium model. Then we made a detailed comparison with the only
full-potential
GW calculation of Hamada {\it et al.} \cite{Hamada} where LAPW-GWA calculation
of the band structure of Si is presented. Table \ref{exchange} 
compares different key
ingredients necessary to evaluate the quasiparticle energies of Si in the GW
approximation. Our calculated matrix elements of the exchange-correlation
potential $\langle\Psi_{{\bf k}n}|V_{xc}^{LDA}[n_{v}({\bf r})]|\Psi_{{\bf
k}n}\rangle$
are compared to those obtained by LAPW-GWA. The agreement with LAPW is
excellent
and reflects the accuracy our LDA results obtained by PAW method. We have
also compared the screened exchange (SEX) and Coulomb-hole (COH) contributions to
the self-energy with the results of Hamada {\it et al.}\cite{Hamada} and of
Hybertsen and Louie\cite{Hybertsen1}.
To make this comparison reliable, we have used the same type of plasmon-pole
model than Hamada {\it et al.} We have found that the agreement of ours results 
with these of Hamada {\it et al.} is not fully satisfactorily, but our results 
are very close to the results of Hybertsen-Louie although their calculation 
is a pseudopotential  one and uses another type of plasmon-pole model. 
This observation
makes us confident with the PAW-GWA results and confirms the fact, 
as outlined
before, that the detailed structure of the screened interaction is not
important to determine the quasiparticle energies. We should point out here
that $\Sigma=\Sigma_{SEX}+\Sigma_{COH}$ which enters the calculation of the
quasiparticle energies are close to each other whether we consider the PAW,
the LAPW, or the Pseudo-Potential implementations.

To calculate the quasiparticle energies it is important to correctly
determine the quasiparticle renormalization factor 
$Z_{n\k}$ (see Eq. \ref{renormalization}). As stated earlier, 
Table \ref{renormalization} presents $Z_{n\k}$ calculated  
for the top valence state at the $\Gamma$ point and for the lowest conduction state 
of all semiconductors studied in this paper. These values
are in good agreement with the results of Hybertsen and Louie
\cite{Hybertsen1} and they seem to be material and state independent and
are at the vicinity of  0.8. 

Tables \ref{si}, \ref{gaas}, \ref{alas}, \ref{inp},  and \ref{mg2si} 
represent the calculated PAW-LDA and PAW-GWA 
band energies for the  high symmetry points  $\Gamma$, $X$, and $L$   
for small and medium band gap semiconductors: Si, GaAs, AlAs, InP, and
Mg$_2$Si, 
respectively (The energy scale is relative to the top of the valence state 
maximum).  These results  are  compared to 
other GWA calculations obtained using  LAPW (for Si) or  PP for the other
systems as well as to available experimental 
data 
\cite{numerical_data,Ortega,Spicer,Wachs,Himpsel1,Hulthen,Straub,Wolford,Aspnes,Cardona,Lautenschlager,Heller}.
In these tables we have presented our results for 2 Chadi-Cohen {\bf
k}-points \cite{chadi_cohen} which correspond to 
32 $\bf k$-points in the whole Brillouin zone (BZ) (results shown between parentheses) 
as well as the 
most converged values using 
10 special {\bf k}-points \cite{Monkhorst} corresponding to 256 $\bf
k$-points in the BZ. 
We notice that most of the data in the literature are produced  
with about 32 $\bf k$-points in the BZ, and are in excellent agreement with our 
unconverged values, however only the converged values should be compared 
to experiment. In our case we have found that the QP eigenvalues are 
converged with a $\bf k$-point mesh beyond 256 points. 
The discrepancy between our GW values and others is often traced to 
differences between LDA values. It's also worth mentioning that 
Shirley {\it et al.}'s
GW results for GaAs\cite{shirleyzhu} agree well with our results 
(see Table \ref{gaas})
despite that they are of PP type, and that their calculation shows that
only the  inclusion of 
core-polarization effect  produce gaps in agreement with  experiment.

\begin{figure}
\begin{minipage}{7.0in}
\epsfxsize=6.4truein
\epsfysize=8.0truein
\centerline{\epsfbox{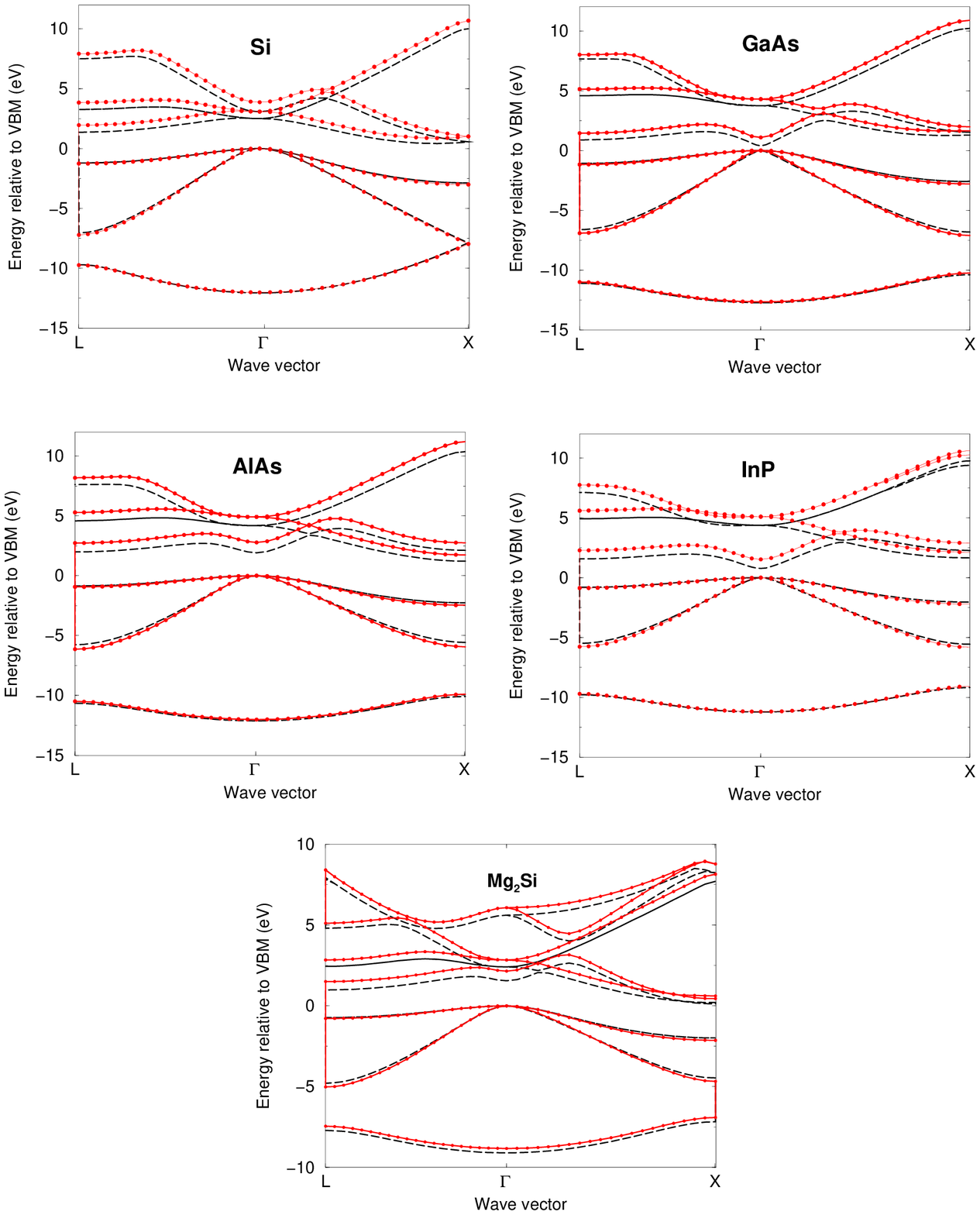}}
\end{minipage}
\caption{
\label{small_band_gaps}
Calculated electronic band structures along  high symmetry directions for 
some small and medium band gap semiconductors:
Si, GaAs, AlAs,  InP, and SiMg$_2$ (in eV). 
The dashed lines display LDA results calculated with an energy
cut-off of 15Ry (cf. Table \ref{si},\ref{gaas},\ref{alas},\ref{inp}, and
\ref{mg2si}, column 2). 
The solid lines with dots show the
GW results based on these LDA results (cf. Tables \ref{si},\ref{gaas},
\ref{alas},\ref{inp}, and  \ref{mg2si}, 
column 3). The energy scale is relative to the top of the valence state
maximum (VBM).
}
\end{figure}

Fig. \ref{small_band_gaps} presents the corresponding band
structure along the $L \Gamma$ and $\Gamma X$ high symmetry directions
calculated within the LDA and GWA. 
We notice an
overall improvement of the  excited states eigenvalues compared to these
obtained in LDA, whereas the LDA valence states eigenvalues are 
already in good agreement with experiment and GWA results 
do not change this agreement. In all these small and medium band gap
semiconductors we remark also  an average energy shift of the conduction states
towards the high energies compared to LDA. 
This energy shift is about the same for Si and GaAs and 
is about 0.6 eV, and increases to about 0.8 eV for AlAs and InP. 
To be more specific we  studied the range of applicability of the 
so called scissors-operator shift. 
To this end  we have evaluated the deviation of  the difference of 
LDA and GWA   direct band gap determined at 
 the $\Gamma$, X, and L symmetry points.
 We have found that these deviation of the 
GWA and LDA energy differences  are the lowest for Si, 
and Mg$_2$Si. The maximum deviation
is  about 0.06, and 0.04 eV, respectively, and it occurs from 
$L$ to $\Gamma$ in both materials. 
The deviations are somewhat larger for GaAs, AlAs, and InP, and the 
maximum deviation is about, 0.16, 0.15, and 0.13 eV, respectively. They are
all   from $\Gamma$ to $X$. 
These small deviations indicate that the GWA does not change much the LDA
dispersion across the Brillouin zone,  justifying  the use of 
the scissors-operator shift 
for the calculation of the dielectric function 
 for small and medium band gap semiconductors \cite{chen,levine,aw}. 
We will see in the next subsection  that such  deviations are  much larger 
in absolute values for wide-band-gap semiconductors.  

As for  Mg$_{2}$Si we believe that it is the first time
that this compound is studied within the GW approximation.
The PAW-LDA  and PAW-GWA band energies for the  high symmetry points  
$\Gamma$, $X$, and $L$ are shown in table \ref{mg2si}. 
Due to the lack of photoemission experiments, the GW results
are compared with optical measurements, making the assumption that 
excitonic effects
are negligible. The GWA results are in good agreement with the experimental
 results and compare favorably with the EPM calculation of Au-Yang et
al.\cite{yangcohen}  Fig. \ref{small_band_gaps} presents the corresponding 
band structure along the $L \Gamma$ and $\Gamma X$ high symmetry directions
within the LDA and GWA.

\begin{figure}
\begin{minipage}{6.0in}
\epsfysize=7.0in
\centerline{\epsfbox{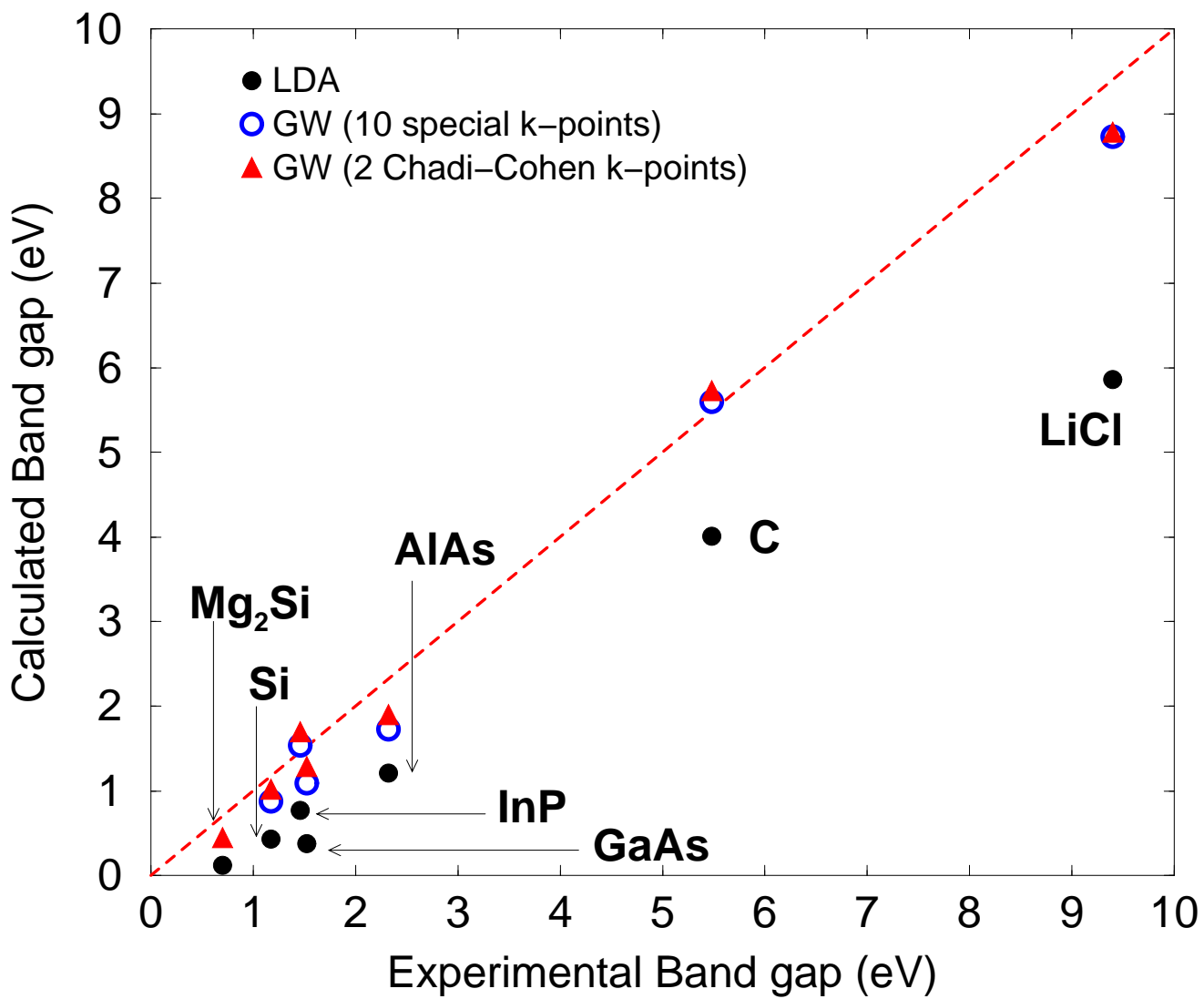}}
\end{minipage}
\caption{
\label{band_gaps}
Calculated LDA and GWA band gap compared to experimental results. 
The filled
circles represent the LDA values, the open circles the GW values with 10 special
{\bf k}-points (corresponding to 256 points in the BZ) and the up-triangles 
the GW values using 2 chadi-Cohen {\bf k}-points (corresponding to 32 points in the BZ).
A perfect agreement with experiment is achieved when the calculated value is 
on the dashed line.
}
\end{figure}

Fig. \ref{band_gaps} shows the LDA and GWA calculated minimum band gaps
for all studied semiconductors, and are 
compared to the experimental results. A perfect agreement with experiment is 
achieved when the calculated value falls 
on the dashed line. We notice that for most of the small and medium gap 
semiconductors GWA does not account for the whole correction 
of the band gap.  The disagreement with experimental band gaps is most
probably due to the  procedure used to  decouple the core and valence 
electrons. 
Indeed, we used the LDA exchange-correlation
potential with the valence electron density to estimate the LDA counterpart
of the self-energy and we believe that a more refined treatment should rely
on the evaluation of the core-valence exchange interaction within the
Hartree-Fock approximation.\cite{Hedin2} 
However, this latter suggested procedure is difficult to
implement in our formalism and the outcome would  not 
necessarily improve upon the former  simple scheme.
On the other hand,  it is interesting to mention that a first order vertex and 
self-consistent corrections to the RPA polarizability and to the self-energy 
within the GWA increase the direct energy band gaps of Si at the 
$\Gamma$, $L$, and $X$ points by about 0.36, 0.44, and 0.39 eV, 
respectively \cite{ummels}. It seems  then that there is no compensation 
between
the vertex correction and the selfconsistency as original thought. 
If we start from our converged Si GWA   results, and use these latter  
results we could improve the agreement between our calculated band gaps
and experimentent.
However, when some  PP-GWA results are used, the additional corrections 
overestimates the experimental gaps. At the present time it looks like 
the question of the band gaps is not fully solved.  It is also important 
to remark  that these vertex corrections and the corrections arising 
from the  selfconsistency are not quite accurate   since the starting point 
of these calculations is a noninteracting Green's
function instead of the selfconsistent Green's function as suggested by
Hedin \cite{Hedin1}.

\subsection{Quasiparticle results for wide band gap semiconductors and 
insulators:  C, and LiCl} 
It is of interest to compare all-electron GWA calculations for wide band
gap semiconductors and insulators to existing PP calculations. Wide band
gap semiconductors are somehow puzzling in contrast to small and medium 
semiconductors: While the LDA band gap of these materials are significantly
underestimated compared to experiment, the LDA static dielectric function
are usually in good agreement with experimental results, see for instant 
Ref. \cite{chen,aw}.

\begin{figure}
\begin{minipage}{6.0in}
\epsfxsize=5.0in
\epsfysize=7.0in
\centerline{\epsfbox{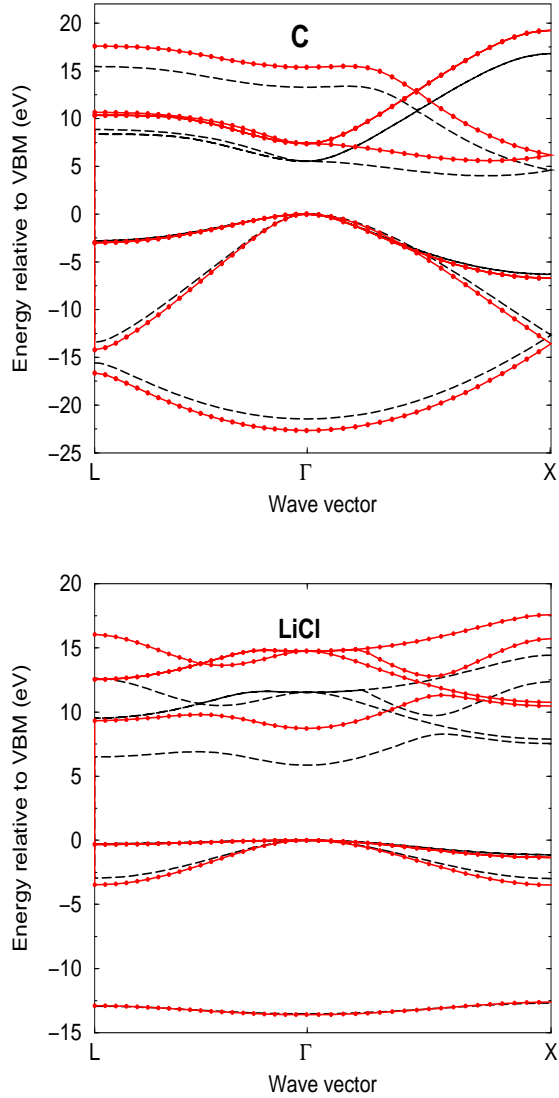}}
\end{minipage}
\caption{
\label{wide_band_gaps}
Calculated electronic band structures along lines of high symmetry for
some large band gap semiconductors: 
C, and LiCl (in eV). 
The dashed lines display LDA results calculated with an energy
cut-off of 45Ry (cf. Table \ref{diamond},\ref{licl}, column 2). The solid lines 
with dots show the
GW results based on these LDA results (cf. Tables \ref{diamond},\ref{licl}, 
column 3). The energy scale is relative to the top of the valence state
maximum (VBM).
}
\end{figure}

Tables \ref{diamond}, and \ref{licl} show the calculated PAW-LDA and PAW-GWA 
band energies for the  high symmetry points  $\Gamma$, $X$, and $L$   
for wide band gap semiconductors: C, and LiCl, 
respectively (The energy scale is relative to the top of the valence state 
maximum).  These results  are  compared to 
other GWA calculations obtained using  PP-GWA method\cite{Hybertsen1,Rohlfing}  
and to  experimental 
data  whenever available\cite{numerical_data,McFeely,Himpsel2,Baldini}. 
Fig. \ref{wide_band_gaps} presents the corresponding band
structure along the $L \Gamma$ and $\Gamma X$ high symmetry directions
within the LDA and GWA. The C and LiCl values  have been computed 
using 32 $\bf k$-points as well as 256 $\bf
k$-points in the Brillouin zone.
The results of the two sets of $\bf k$-points
are in good agreement showing that the set of 32 $\bf k$-points is good
enough. 
For C the calculated QP eigenvalues are in good agreement with
experiment and the PP calculations. For LiCl only the experimental
band gap is available  and is slightly larger than our GW value.
It's worth mentioning that we didn't update the Green function to get our
GW values. Such a procedure leads to an increase of the GW band gap by
about 0.3 eV and then to a better agreement between our results and 
PP results of Hybertsen and Louie
\cite{Hybertsen1}.
For these wide-gap materials we looked also to the 
 applicability of the scissors-operator shift. We calculated the 
maximum change of the difference between the GWA and LDA  direct band gap 
at the $\Gamma$, X, and L symmetry points. We found that the maximum
deviation across the Brillouin zone is for C  and is about 0.32 eV. 
It occurs  for the  L and $\Gamma$ differences,whereas 
it is about 0.28 eV for LiCl and occurs for  X and $\Gamma$ energy differences. 
These  deviations seem to be  somewhat larger in absolute values 
(about twice the value found for GaAs) than for small and medium-gap 
semiconductors,
and may make the  use of the  scissors-operator shift, 
for the computation  of the optical properties, less applicable. 
However, 
if we compare  these energy deviations  to the size of 
the band gap, we find that the largest ratio occurs for C and is only 5\%, 
whereas it is about 10\% for a medium gap semiconductor such as GaAs.

\section{Conclusion}
We  have  implemented a  GWA based on an all-electron method using the recently 
developed projector augmented wave (PAW) method.\cite{Blochl}
The knowledge of the one-electron Green function of the PAW Hamiltonian 
allows us to construct the quasiparticle self-energy within the GWA,
in which the dynamical screening of the electron-electron
interaction arises from a plasmon model  dielectric
function\cite{Horsch,Hamada,Engel}  for which
the parameters are adjusted to the dielectric function calculated using the
random-phase-approximation (RPA).  
We have tried various plasmon model  dielectric functions for the screening
of the Coulomb interaction and showed that the quasiparticle energies are
insensitive to the type of the model used. 

Using this new GWA method, we have determined for the first time the 
GWA quasiparticle electronic
structure of Mg$_2$Si. We have found that our LDA results are in good
agreement with the empirical pseudopotential, and that the GWA shifts almost
rigidly the empty states by about 0.32 eV towards higher energies. 

Concerning the other semiconductors studied here, we have found an overall 
agreement of  our calculated electronic structure of various
semiconductors with existing GWA pseudopotential calculations performed by 
different groups\cite{Hybertsen1,Godby3,Hott,shirleyzhu,Rohlfing},  and with 
experimental results
\cite{numerical_data,McFeely,Himpsel2,Baldini,Ortega,Spicer,Wachs,Himpsel1,Hulthen,Straub,Aspnes,Cardona,Lautenschlager,Heller,Wolford}.
However, for detailed comparisons, the all-electron GWA band gaps are
slightly smaller that most of  the  PP results. Nevertheless,  some of the PP
results\cite{shirleyzhu} are much closer to ours.
One of the possible source of the  the all-electron results compared to the PP
ones is the way the  decoupling of  
the core and valence electrons is done. In the PP approach  this decoupling is 
some what  hidden.  In fact, we have used  
the LDA exchange-correlation potential with the valence electron density to 
remove  the LDA counterpart of the self-energy.  We believe that a more 
refined treatment should rely
on the evaluation of the core-valence exchange interaction within the
Hartree-Fock approximation\cite{Hedin2} which we find difficult to implement 
in our current  all-electon PAW-GWA approach. 
On the other hand,  the first order vertex and 
self-consistent corrections  to the RPA polarizability and to the self-energy 
within the GWA are shown to increase the direct energy band gaps of Si at the 
$\Gamma$, $L$, and $X$ points by few tenths of an eV\cite{ummels}. 
It seems  then that there is no compensation 
between the vertex correction and the selfconsistency as it was always
assumed. However, these corrections are not quite accurate since the  
starting point of these calculations is a noninteracting Green's
function instead of the selfconsistent Green's function as suggested by
Hedin \cite{Hedin1}.

 To our knowledge this is the first all-electron 
GWA calculation that has corrected   LDA eigenvalues for
three type of semiconductors: small, medium, and wide band gap,
and that questioned the correctness of the band gap of
semiconductors obtained by means of  PP-GWA. 
We hope that this work would  be used as a reference and triggers off  
 further interest on  an all-electron GW  approach.

\section{Acknowledgment}
We would like to thank P. Bl\"ochl for providing us with his PAW code 
and for useful discussions. Part of this work was done during our
visit to the Ohio State University, and we would like to thank 
J. W. Wilkins and W. Aulbur for useful discussions.  
The Supercomputer time was granted by CINES
on the IBM SP2 supercomputer  (project gem1100). 

%

\begin{table}
\caption{The renormalization constants $Z$ for the hole state
at the top of the valence band (VBM) and the electron state
near the bottom of the conduction band (CBM) for C, Si,
GaAs, AlAs, InP and SiMg$_2$.}
\label{renormalization}
\begin{tabular}{ccccccc}
                                & C & Si     &GaAs & AlAs & InP & Mg$_2$Si \\ \hline
Z$_{{\textrm VBM }}$             &0.85    & 0.80   &0.80  &0.81 &0.79 &0.76   \\
Z$_{{\textrm CBM }}$             &0.87    & 0.81   &0.81  &0.82 &0.82 &0.79   \\

\end{tabular}
\end{table}

\newpage

%

\begin{table}
\caption{Quasiparticle (QP) energies of Si for several states (in eV)
and for three different types of plasmon-pole models. We notice that 
the QP energies are less sensitive to the type of plasmon pole model used.}
\label{qp_si_3_plasmon}
\begin{tabular}{lccc}
                                &\multicolumn{3}{c}{Plasmon-pole model}  \\
                                & Von der Linden and Horsch$^a$ & Hamada
{\it et al}$^b$   &
Engel-Farid$^c$ \\
 \hline
${\bf{\Gamma}}_{1v}$            & -12.01  & -11.99  &-11.89       \\
${\bf{\Gamma}}_{25^{\prime}v}$  & 0.00    & 0.00    & 0.00        \\
${\bf{\Gamma}}_{15c}$           & 3.09    & 3.07    & 3.12         \\
${\bf{\Gamma}}_{2^{\prime}c}$   & 3.88    & 3.86    & 3.86         \\
                                &         &         &              \\
${\bf X}_{1v}$                  & -7.97   & -7.96   &-8.01         \\
${\bf X}_{4v}$                 & -3.00    & -3.00   &-3.01         \\
${\bf X}_{1c}$                 &  1.01    & 0.99    &1.04          \\
                               &          &         &             \\
${\bf L}_{2^{\prime}v}$         & -9.73   & -9.72   &-9.75         \\
${\bf L}_{1v}$                  & -7.21   & -7.20   &-7.19          \\
${\bf L}_{3^{\prime}v}$         & -1.25   & -1.25   &-1.29           \\
${\bf L}_{1c}$                  & 1.96    & 1.94    &1.94            \\
${\bf L}_{1c}$                  & 3.85    & 3.83    &3.86            \\
\end{tabular}
$^a$Ref.\cite{Horsch}, $^b$Ref.\cite{Hamada}, $^c$Ref.\cite{Engel}
\end{table}

\newpage
%
\begin{table}
\caption{Matrix elements of the exchange-correlation potential
and of the screened exchange (SEX) and the Coulomb hole (COH) 
contributions to the self-energy
of Si for several states (in eV). The $V_{XC}$ is in excellent agreement
with the LAPW results\cite{Hamada}, whereas the SEX and COH are much
closer to the PP work of Hybertsen and Louie \cite{Hybertsen1} and disagree
with the LAPW results.}
\label{exchange}
\begin{tabular}{lcccccccc}
                                & \multicolumn{2}{c}{$V_{XC}$} &
\multicolumn{3}{c}{$\Sigma_{SEX}$}
& \multicolumn{3}{c}{$\Sigma_{COH}$} \\
                                & Present   & LAPW$^a$              &Present
& LAPW    & H-L$^b$
& Present & LAPW  & H-L  \\ \hline
${\bf{\Gamma}}_{1v}$            & -10.59    & -10.60                &-3.81
& -4.44   &
& -7.46   & -6.97 &       \\
${\bf{\Gamma}}_{25^{\prime}v}$  & -11.44    & -11.41                &-3.44
& -4.01   & -3.56
& -8.66   & -7.80 & -8.41  \\
${\bf{\Gamma}}_{15c}$           & -10.17    & -10.19                &-1.80
& -1.95   &
& -8.32   & -7.68 &         \\
${\bf{\Gamma}}_{2^{\prime}c}$    & -11.29   & -11.25                &-1.59
& -1.90   &
& -9.30   & -8.43 &         \\
                                &           &                       &
&         &
&         &       &       \\
${\bf X}_{1v}$                 & -10.97     & -10.98                &-3.92
& -4.56   &
& -7.90   & -7.23 &          \\
${\bf X}_{4v}$                 & -10.74     & -10.73                &-3.49
& -3.89   &
& -8.10   & -7.48 &           \\
${\bf X}_{1c}$                 & -9.15      & -9.17                 &-1.76
& -1.85   & -1.65
& -7.50   & -7.09 & -7.40      \\
                               &            &                       &
&         &
&         &       &             \\
${\bf L}_{2^{\prime}v}$         & -10.97    & -10.97                &-3.97
& -4.70   &
& -7.78   & -7.11 &             \\
${\bf L}_{1v}$                  & -10.35    & -10.37                &-3.65
& -3.92   &
& -7.64   & -7.29 &              \\
${\bf L}_{3^{\prime}v}$         & -11.20    & -11.18                &-3.47
& -4.02   &
& -8.46   & -7.65 &              \\
${\bf L}_{1c}$                  & -10.28    & -10.30                &-1.87
& -2.08   &
& -8.34   & -7.68 &               \\
${\bf L}_{1c}$                  & -9.77     & -9.77                 &-1.64
& -1.81   &
& -8.08   & -7.43 &               \\
\end{tabular}
$^a$Ref.\cite{Hamada}, $^b$Ref.\cite{Hybertsen1}
\end{table}

\newpage

%

\begin{table}
\caption{Quasiparticle energies
of Si for several states (in eV). The calculation of the self-energy is performed using
10 special ${\bf k}$-points, 200 bands and 283 reciprocal lattice vectors. The size of
the polarizability matrix is 137$\times$137 and the plasmon pole model of von der 
Linden and Horsch \protect{\cite{Horsch}}  is
used.  Here $E_g$ is the minimum band gap.}
\label{si}
\begin{tabular}{lcccccc}
                                & \multicolumn{2}{c}{LDA}       & \multicolumn{3}{c}{GW approximation}     & Expt.$^a$ \\ 
                                & Present   &LAPW$^b$      &   Present            &  LAPW$^b$   &  (HL)$^c$    &          \\ \hline
${\bf{\Gamma}}_{1v}$            & -12.05    & -11.95       & -12.01 (-11.99)      &  -12.21 &  -12.04      & -12.5$\pm$0.6\\
${\bf{\Gamma}}^{\prime}_{25v}$  & 0.00      & 0.00         &   0.00 (0.00)        &    0.00 &    0.00      &   0.00 \\
${\bf{\Gamma}}_{15c}$           & 2.51      & 2.55         &   3.09 (3.23)        &    3.30 &    3.35      &   3.40,3.05$^d$ \\
${\bf{\Gamma}}^{\prime}_{2c}$   & 3.10      & 3.17         &   3.88 (4.02)        &    4.19 &    4.08      &   4.23, 4.1$^d$   \\
                                &           &              &             &         &              &                    \\
${\bf X}_{1v}$                  & -7.88     & -7.82        &   -7.97 (-8.07)        &   -8.11 &              &                     \\
${\bf X}_{4v}$                  & -2.87     & -2.84        &   -3.00 (-3.04)    &   -3.03 &    -2.99     &   -2.9$^e$, -3.3$\pm$0.2$^f$\\
${\bf X}_{1c}$                  &  0.56     & 0.65         &    1.01 (1.12)     &    1.14 &     1.44     &    1.25$^d$\\
${\bf X}_{4c}$                  & 10.01     &              &   10.69 (10.77)    &         &              &            \\
                                &           &              &         &              &             \\
${\bf L}^{\prime}_{2v}$         & -9.70     & -9.63        &   -9.73 (-9.81)    &   -9.92 &    -9.79     &   -9.3$\pm$0.4\\
${\bf L}_{1v}$                  & -7.04     & -6.98        &   -7.21 (-7.26)    &   -7.31 &    -7.18     &   -6.7$\pm$0.2  \\
${\bf L}^{\prime}_{3v}$         & -1.20     & -1.19        &   -1.25 (-1.29)    &   -1.26 &    -1.27     &   -1.2$\pm$0.2,-1.5$^g$  \\
${\bf L}_{1c}$                  &  1.37     & 1.43         &    1.96 (2.06)     &    2.15 &     2.27     &    2.1$^h$,2.4$\pm$0.15  \\
${\bf L}_{3c}$                  &  3.27     & 3.35         &    3.85 (3.96)     &    4.08 &     4.24     &    4.15$\pm$0.1$^g$ \\
$E_g$ gap                     &  0.43     & 0.52         &    0.88 (1.03)     &    1.01 &     1.29     &    1.17 \\ 
\end{tabular}
$^a$Ref.{\cite{numerical_data}}, $^b$Ref.\cite{Hamada}, $^c$Ref.\cite{Hybertsen1}, 
$^d$Ref.\cite{Ortega}, $^e$Ref.\cite{Spicer}, $^f$Ref.\cite{Wachs}, 
$^g$Ref.\cite{Himpsel1}, $^h$Ref.\cite{Hulthen}, $^i$Ref.\cite{Straub}.  
\end{table}

%

\begin{table}
\caption{Quasiparticle energies
of GaAs for several states (in eV). The calculation of the self-energy is performed using
10 special ${\bf k}$-points, 200 bands and 307 reciprocal lattice vectors. The size of
the polarizability matrix is 169$\times$169 and the plasmon pole model of von der Linden
and Horsch \protect{\cite{Horsch}} is
used. Our results are compared with those of
Rohlfing {\emph et al}\cite{Rohlfing} and those of Shirley {\emph et al}
\cite{shirleyzhu}. The results of Shirley are given without (with) core
polarization effects. Here $E_g$ is the minimum band gap.} 
\label{gaas}
\begin{tabular}{lcccccc}
                        & \multicolumn{2}{c}{LDA}       & \multicolumn{3}{c}{GW approximation}                   & Expt.$^a$ \\
                        &  Present  &  (RKP)$^b$        &   Present            &  (RKP)$^b$   &  (SZL)$^c$       &          \\ \hline
${\bf{\Gamma}}_{1v}$    & -12.71    & -12.69            & -12.64 (-12.62)      &  -12.69      &                  & -13.21\\
${\bf{\Gamma}}_{15v}$   & 0.00      & 0.00              &   0.00 (0.00)        &    0.00      &    0.0           &   0.00 \\
${\bf{\Gamma}}_{1c}$    & 0.38      & 0.57              &   1.09 (1.29)        &    1.32      &    1.02, (1.42)  &   1.5 \\
${\bf{\Gamma}}_{15c}$   & 3.74      & 3.73              &   4.30 (4.46)        &    4.60      &                  &  4.61   \\
                        &           &                   &                      &              &                  &        \\
${\bf X}_{1v}$          &-10.37     &-10.37             &   -10.23 (-10.38)    &  -10.27      &                  & -10.86    \\
${\bf X}_{3v}$          &-6.81      &-6.79              &   -7.10  (-7.22)     &   -7.16      &                  & -6.81  \\
${\bf X}_{5v}$          &-2.59      &-2.56              &   -2.79  (-2.83)     &   -2.71      &                  & -2.91  \\
${\bf X}_{1c}$          & 1.29      &1.80               &    1.64  (1.75)      &   2.65       &     2.07, (1.95) & 1.90 \\
${\bf X}_{3c}$          & 1.53      &1.85               &    1.98 (2.09)       &   2.72       &                  &      2.47      \\
${\bf X}_{5c}$          & 10.20     &10.33              &    10.88 (10.95)     &  11.20       &                  &           \\
                        &           &                   &                      &              &                  &             \\
${\bf L}_{1v}$          & -11.09    &-11.08             &   -10.99 (-11.08)    &  -11.02      &                  &   -11.35 \\
${\bf L}_{1v}$          & -6.61     &-6.59              &   -6.91 (-6.97)      &  -6.91       &                  &   -6.81  \\
${\bf L}_{3v}$          & -1.08     &-1.10              &   -1.17 (-1.22)      &  -1.17       &                  &     -1.41 \\
${\bf L}_{1c}$          &  0.89     &1.13               &    1.45 (1.57)       &  1.92        &    1.55, (1.75)  &     1.74 \\
${\bf L}_{3c}$          &  4.58     &4.67               &    5.12 (5.25)       &  5.65        &                  &   5.45$^d$  \\
${\bf L}_{1c}$          &  7.65     &8.88               &    8.13 (8.00)       & 9.92         &                  &   8.6$^d$       \\
$E_g$             &  0.38     &0.57               &    1.09 (1.29)       & 1.32         & 1.02,
(1.42)                 &    1.5 \\
\end{tabular}
$^a$Ref.{\cite{numerical_data}}, $^b$Ref.\cite{Rohlfing}, $^c$Ref.\cite{shirleyzhu},
$^d$Ref.\cite{Ortega}
\end{table}

%

\begin{table}
\caption{Quasiparticle energies
of AlAs for several states (in eV). The calculation of the self-energy is performed using
10 special ${\bf k}$-points, 200 bands and 283 reciprocal lattice vectors. The size of
the polarizability matrix is 169$\times$169 and the plasmon pole model of von der Linden and
Horsch \protect{\cite{Horsch}} is
used. Our results are compared with those of
Godby {\emph et al}\cite{Godby2} (GS) obtained using the RPA dielectric function for the screening of
the Coulomb interaction, and 
these of  Shirley, Zhu, and Louie (SZL) using a plsmon-pole model.
The Godby  {\emph et al}   results include spin-orbit coupling and
the lower energy of a spin orbit pair is shown between brackets. 
Here $E_g$ is the minimum band gap. } 
\label{alas}
\begin{tabular}{lcccccc}
                                & \multicolumn{2}{c}{LDA}       & \multicolumn{3}{c}{GW approximation}      & Expt.$^a$ \\
                                & Present   & GS$^b$          &   Present          &  GS$^b$ &SZL$^e$&             \\ \hline
${\bf{\Gamma}}_{1v}$            & -12.13    &                   & -12.01 (-11.99)   &        &               &   \\
${\bf{\Gamma}}^{\prime}_{15v}$  & 0.00      & 0.00 (-0.28)      &  0.00  (0.00)     &    0.00 [-0.28]  &0.00 & 0.00 \\
 ${\bf{\Gamma}}_{1c}$            & 1.92      & 2.29              &   2.79  (2.97)     &     3.26 & 2.75   &3.11$^c$    \\
${\bf{\Gamma}}_{15c}$           & 4.19      & 4.23              &   4.91  (5.07)     &    5.05 &          &     \\
                                &           &                   &                    &       &            &           \\
${\bf X}_{1v}$                  &-10.08     &                   &   -9.90 (-9.88)    &       &            &      \\
${\bf X}_{3v}$                  &-5.58      &                   &   -5.94 (-5.93)    &       &                 &     \\
${\bf X}_{5v}$                  &-2.26      & -2.21 (-2.36)     &   -2.46 (-2.45)    &   -2.34 [-2.49] &    &-2.30 \\
${\bf X}_{1c}$                  & 1.21      & 1.28              &    1.73 (1.90)     &   2.09 & 2.09       &  2.24    \\
${\bf X}_{3c}$                  & 2.12      & 2.14              &    2.75 (2.90)     &   2.99&               &        \\
${\bf X}_{5c}$                  & 10.38     &                   &   11.22 (11.36)    &  11.20&               &       \\
                                &           &                   &                    &   &                   &           \\
${\bf L}_{1v}$                  & -10.65    &                   &   -10.50 (-10.48)  &   &                   &     \\
${\bf L}_{1v}$                  & -5.76     &                   &  -6.14  (-6.12)    &   &                   &        \\
${\bf L}_{3v}$                  & -0.85     & -0.80 (-1.00)     &   -0.93  (-0.92)   & -0.88 [-1.08]&        & -1.31   \\
${\bf L}_{1c}$                  &  2.00     & 2.13              &    2.73  (2.89)    &  3.03 & 2.81  &  2.49$^c$, 2.54$^d$\\
${\bf L}_{3c}$                  &  4.59     & 4.58              &    5.29  (5.45)    &  5.48&                &        \\
${\bf L}_{1c}$                  &  7.62     &                   &    8.19  (8.34)    &    &                  &          \\
$E_g$                     &  1.21     & 1.37              &    1.73  (1.90)    &  2.09 & 2.09               &    2.32    \\     
$\Gamma$ direct gap             &  1.92     & 2.29              &    2.79 (2.97)     & 3.26  & 2.75       &    3.11$^c$  \\
$X$ direct gap                  &  3.47     & 3.47              &    4.19 (4.35)     & 4.41 &             &    4.54 \\
$L$ direct gap                  &  2.85     & 2.93              &    3.66 (3.81)     & 3.91  &           &    3.90$^d$  \\
\end{tabular}
$^a$Ref.{\cite{numerical_data}}, $^b$Ref.\cite{Godby2}, $^c$Ref.\cite{Wolford},
$^d$Ref.\cite{Aspnes}, $^e$ Ref.\cite{shirleyzhu}
\end{table}

%

\begin{table}
\caption{Quasiparticle energies
of InP for several states (in eV). The calculation of the self-energy is performed using
10 special ${\bf k}$-points, 200 bands and 331 reciprocal lattice vectors. The size of
the polarizability matrix is 169$\times$169 and the plasmon pole model of von der Linden
and Horsch \protect{\cite{Horsch}} is
used. Our results are compared with those of
Hott\cite{Hott}.  Here $E_g$ is the minimum band gap.}
\label{inp}
\begin{tabular}{lccccc}
                                & \multicolumn{2}{c}{LDA} & \multicolumn{2}{c}{GW approximation}     & Expt.$^a$ \\
                                &  Present  & (H)$^b$     &   Present            &  (H)$^b$                   &     \\ \hline
${\bf{\Gamma}}_{1v}$            & -11.22    & -11.50       & -11.20 (-11.18)     &   -11.75                   & -11.6   \\
${\bf{\Gamma}}^{\prime}_{15v}$  & 0.00      & 0.00         &   0.00 (0.00)       &    0.00                   &0.00     \\
${\bf{\Gamma}}_{1c}$            & 0.77      & 0.50         &   1.54 (1.70)       &     1.23                  &1.460$^c$    \\
${\bf{\Gamma}}_{15c}$           & 4.37      & 4.21         &   5.10 (5.24)       &    5.17                   & 5.00$^d$      \\
                                &           &              &                     &                           &   \\
${\bf X}_{1v}$                  &-9.18      & -9.29        &   -9.08 (-9.14)     &  -9.16                    &   -9.24   \\
${\bf X}_{3v}$                  &-5.56      & -5.94        &   -5.82 (-5.84)     & -6.60                     &   -5.93  \\
${\bf X}_{5v}$                  &-2.04      & -2.34        &   -2.21 (-2.21)     &  -2.52                    &-2.40 \\
${\bf X}_{1c}$                  & 1.66      & 1.64         &    2.13 (2.26)      &   2.60                    & 2.42    \\
${\bf X}_{3c}$                  & 2.28      & 2.10         &    2.90 (3.02)      &   2.63                    & 2.92\\
${\bf X}_{5c}$                  & 9.39      & 9.45         &   10.26 (10.35)     &  10.94                    &       \\
                                &           &              &                     &                           &           \\
${\bf L}_{1v}$                  & -9.78     & -9.94        &   -9.71 (-9.74)     &  -9.97                    &     -9.89\\
${\bf L}_{1v}$                  & -5.48     & -5.90        &   -5.77 (-5.77)     &  -6.56                    &     -5.93   \\
${\bf L}_{3v}$                  & -0.81     & -0.94        &   -0.87 (-0.88)     & -0.94                     & -0.98   \\
${\bf L}_{1c}$                  &  1.57     & 1.30         &    2.28 (2.41)      &  1.97                     &  2.32     \\
${\bf L}_{3c}$                  &  4.92     & 4.75         &    5.60 (5.72)      &  5.85                     &  5.68      \\
${\bf L}_{1c}$                  &  7.12     & 7.47         &    7.74 (7.86)      &  8.17                     &            \\
$E_g$                     &  0.77     & 0.50         &    1.54 (1.70)      & 1.23                      &    1.460$^c$  \\
\end{tabular}
$^a$Ref.{\cite{Cardona}}, $^b$Ref.\cite{Hott}, $^c$Ref.\cite{Godby3},
$^d$Ref.\cite{Lautenschlager}
\end{table}

%
\begin{table}
\caption{Quasiparticle energies
of Mg$_{2}$Si for several states (in eV). The calculation of the
self-energy is
performed using
2 special ${\bf k}$-points, 200 bands and 645 reciprocal lattice
vectors. The size of
the polarizability matrix is 113$\times$113 and the plasmon pole model
of von der Linden and Horsch\protect{\cite{Horsch}} is used. Here $E_g$ is the minimum band gap. }
\label{mg2si}
\begin{tabular}{lccc}
                                                 & LDA       & GW
approximation
 & Expt.$^a$\\ \hline
${\bf{\Gamma}}_{1v}$                             & -9.19     & -8.82
 &            \\
${\bf{\Gamma}}_{15v}$                            & 0.00      & 0.00
 &   0.00      \\
${\bf{\Gamma}}_{1c}$                             & 1.55      & 2.15
 &   2.1        \\
${\bf{\Gamma}}^{\prime}_{25c}$                   & 2.41      & 2.84
 &               \\
                                                 &           &
 &                \\
${\bf X}_{1v}$                                   &-7.17      & -6.91
 &                    \\
${\bf X}^{\prime}_{4v}$                          &-4.46      & -4.67
 &                     \\
${\bf X}^{\prime}_{5v}$                          &-1.99      & -2.14
 &         \\
${\bf X}_{3c}$                                   &0.12       & 0.45
 &          \\
${\bf X}_{1c}$                                   &0.20       & 0.62
 &          \\
                                                 &           &
 &           \\
${\bf L}_{1v}$                                   &-7.71      & -7.45
 &           \\
${\bf L}^{\prime}_{2v}$                          &-4.79      & -5.02
 &            \\
${\bf L}^{\prime}_{3v}$                          &-0.73      & -0.78
 &            \\
${\bf L}_{1c}$                                   &0.98       & 1.50
 &           \\
${\bf L}_{3c}$                                   &2.44       & 2.84
 &           \\
                                                 &           &
 &                \\
${\bf L}^{\prime}_{3v}\rightarrow {\bf L}_{3c}$  &3.17       & 3.62
 &   3.7          \\
${\bf X}^{\prime}_{5v}\rightarrow {\bf X}_{1c}$  &2.19       & 2.76
 &   2.5          \\
$E_g$                                      &0.12       & 0.45
 &  0.7-0.80   \\
\end{tabular}
$^a$Ref.{\cite{Heller}}
\end{table}

%

\begin{table}
\caption{Quasiparticle energies
of C for several states (in eV). The calculation of the self-energy is performed using
10 special ${\bf k}$-points, 200 bands and 387 reciprocal lattice vectors. The size of
the polarizability matrix is 169$\times$169 and the plasmon pole model of von der Linden
and Horsch \protect{\cite{Horsch}} is used.  Here $E_g$ is the minimum band gap.}
\label{diamond}
\begin{tabular}{lcccccc}
                                & \multicolumn{2}{c}{LDA}       & \multicolumn{3}{c}{GW approximation}     & Expt.$^a$ \\
                                & Present   & (RKP)$^b$          &   Present   &  (RKP)$^b$   &  (HL)$^c$    &          \\ \hline
${\bf{\Gamma}}_{1v}$            & -21.45    & -21.35      & -22.66 (-22.65)      &  -22.88 &  -23.0      & -24.2$\pm$1$^d$, -21$\pm$1$^e$\\
${\bf{\Gamma}}^{\prime}_{25v}$  & 0.00      & 0.00        &   0.00 (0.00)     &    0.00 &    0.0      &   0.00 \\
${\bf{\Gamma}}_{15c}$           & 5.53      & 5.58        &   7.39 (7.51)     &    7.63 &    7.5      &   7.3 \\
${\bf{\Gamma}}^{\prime}_{2c}$   &13.29      & 13.10       &   15.38 (15.50)     &   14.54 &   14.8       &  15.3$\pm$0.5$^e$   \\
                                &           &             &             &         &              &                    \\
${\bf X}_{1v}$                  &-12.68     & -12.61      &   -13.58 (-13.56)     &  -13.80 &              &                     \\
${\bf X}_{4v}$                  &-6.30      & -6.26       &   -6.72 (-6.71)       &   -6.69 &         &   \\
${\bf X}_{1c}$                  & 4.61     &  4.63        &   6.19 (6.32)     &    6.30 &          &  \\
${\bf X}_{4c}$                  & 16.80     & 16.91       &   19.26 (19.37)    &   19.50 &              &            \\
                                &           &             &             &         &              &             \\
${\bf L}^{\prime}_{2v}$         & -15.58    & -15.51      &   -16.66 (-16.65)   &  -16.95 &    -17.3     &   -15.2$\pm$0.3$^e$ \\
${\bf L}_{1v}$                  & -13.40    & -13.33      &   -14.20 (-14.19)  &  -14.27 &    -14.4     &   -12.8$\pm$0.3$^e$  \\
${\bf L}^{\prime}_{3v}$         & -2.79     & -2.78       &  -2.99   (-2.98)  &  -2.98  &                &     \\
${\bf L}_{1c}$                  &  8.38     & 8.39        &   10.36  (10.48)  &  10.23 &               &      \\
${\bf L}_{3c}$                  &  8.86     & 8.76        &    10.66 (10.77)   &  10.63 &               &     \\
${\bf L}^{\prime}_{2c}$         &  15.45    & 15.67       &    17.59 (17.71)   & 18.14  &     17.9      &    20$\pm$1.5$^e$        \\
$E_g$                     &  4.01     & 4.01        &    5.60 (5.73)    & 5.67   &     5.6       &    5.48 \\
\end{tabular}
$^a$Ref.{\cite{numerical_data}}, $^b$Ref.\cite{Rohlfing}, $^c$Ref.\cite{Hybertsen1},
$^d$Ref.\cite{McFeely}, $^e$Ref.\cite{Himpsel2}
\end{table}

%

\begin{table}
\caption{Quasiparticle energies
of LiCl for several states (in eV). The calculation of the self-energy is performed using
10 special ${\bf k}$-points, 200 bands and 331 reciprocal lattice vectors. The size of
the polarizability matrix is 259$\times$259 and the plasmon pole model of von der Linden
and Horsch \protect{\cite{Horsch}} is
used. Here $E_g$ is the minimum band gap. }
\label{licl}
\begin{tabular}{lccccc}
                                & \multicolumn{2}{c}{LDA}       & \multicolumn{2}{c}{GW approximation}   & Expt.$^a$\\
                                & Present   & (HL)$^b$          &   Present         &    (HL)$^b$              &          \\ \hline
${\bf{\Gamma}}_{15v}$           & 0.00      & 0.00              &   0.00 (0.00)     &    0.00                  &           \\
${\bf{\Gamma}}_{1c}$            & 5.86      & 6.00              &   8.73 (8.78)     &     9.1                  &           \\
${\bf{\Gamma}}^{\prime}_{25c}$  & 11.54     & 11.8              &   14.75 (14.79)   &   14.91                  &     \\
                                &           &                   &                   &                          &      \\
${\bf X}^{\prime}_{4v}$         &-2.98      & -3.0              &  -3.49 (-3.47)    &   -3.3                   &      \\
${\bf X}^{\prime}_{5v}$         &-1.13      &  -1.1             &  -1.34 (-1.33)    &   -1.3                   &       \\
${\bf X}_{1c}$                  & 7.54      & 7.5               &  10.48 (10.52)    &   10.7                   &        \\
${\bf X}_{3c}$                  & 7.89      & 8.2               &  10.77 (10.80)    &   11.6                   &         \\
                                &           &                   &                   &                          &          \\
${\bf L}^{\prime}_{2v}$         & -2.95     & -2.9              &  -3.47 (-3.46)    &  -3.2                    &           \\
${\bf L}^{\prime}_{3v}$         & -0.28     & -0.2              &  -0.32 (-0.31)    &  0.3                     &           \\
${\bf L}_{1c}$                  &  6.50     & 6.4               &   9.33 (9.37)     &  9.7                     &           \\
${\bf L}_{3c}$                  &  9.52     & 9.03              &   12.55 (12.58)   &  12.5                    &            \\
$E_g$                     &  5.86     & 6.00              &   8.73  (8.78)    &  9.1                     &   9.4   \\
\end{tabular}
$^a$Ref.{\cite{Baldini}}, $^b$Ref.\cite{Hybertsen1}
\end{table}

%

%

\newpage

%
\newpage

%
%
\newpage

%

\end{document}